\documentclass[12pt%,draft
]{article}%%%%%% Choose Draft version or final
\usepackage[british]{babel}
\def\AnswerYes{y}
\def\pdflatex{y}                  %%%%%% use pdflatex? (if not, latex)
\def\ShowLabelsVersion{n}         %%%%%% Version with defs. of refs, cites or final
\def\ShowChangesVersion{n}        %%%%%% Version with changes highlighted or final
\def\ShowAnnotationsVersion{n}    %%%%%% Version with annotations or final

\def\feynVersion{n}                %%%%%% Choose whether feynman graphs to be
\ifx\pdflatex\AnswerYes
   \usepackage[update,prepend]{epstopdf} % automaticlaly converts .eps to .pdf file
   \usepackage{grffile} % allows graphics files to have "." in filename
\fi
\ifx\pdflatex\AnswerYes
    \usepackage{color}
\else
    \usepackage[dvips]{color}
\fi
\ifx\pdflatex\AnswerYes
    \usepackage{thumbpdf}    %
\else
    \usepackage[dvips]{thumbpdf}
\fi
   \usepackage{xparse}
   \NewDocumentCommand{\arxiv} %
   {r [: u{ [} u{]]} }{[\href{http://arxiv.org/abs/#2}{arXiv:#2}~[#3]]}
   \NewDocumentCommand{\arxivold} {r[]}{[\href{http://arxiv.org/abs/#1}{#1}]}
   \NewDocumentCommand{\arXiv} %
   {r [: u{ [} u{]]} }{[\href{http://arxiv.org/abs/#2}{arXiv:#2}~[#3]]}
   \NewDocumentCommand{\arXivold} {r[]}{[\href{http://arxiv.org/abs/#1}{#1}]}
\ifx\pdflatex\AnswerYes
    \usepackage[ 
        final,
        breaklinks=true,
        colorlinks=true,           % coloured links 
        pdfborder={0 0 1},    % coloured box about link
        citecolor={blue},linkcolor={blue},urlcolor={red},
        citebordercolor={1 0 0},linkbordercolor={0 0 1},urlbordercolor={1 0 0},
        pdfpagemode=UseOutlines,% whether Acrobat opens with
        bookmarks=true,bookmarksopenlevel=4% write .pdf table of contents.
        ]{hyperref}         % Hyperlinks for PDF versions
\else
    \usepackage[dvips,              % dvips,ps2pdf,pdftex driver 
        final,
        breaklinks=true,
        colorlinks=true,      % coloured links 
        pdfborder={0 0 1},    % coloured box about link
        citecolor={blue},linkcolor={blue},urlcolor={red},
        citebordercolor={1 0 0},linkbordercolor={0 0 1},urlbordercolor={1 0 0},
        pdfpagemode=UseOutlines,% whether Acrobat opens with
        bookmarks=true,bookmarksopenlevel=4% write .pdf table of contents.
         ]{hyperref}         % Hyperlinks for PDF versions
\fi
\usepackage[numbers,sort&compress]{natbib}%%% collapses [1,2,3] -> [1-3] etc
\usepackage[final]{graphicx}            %%% for pictures and figures
\graphicspath{{figures/}{./}}

\usepackage{multirow}                   %%% tabs over multiple rows
\usepackage{amssymb}
\usepackage{amsmath}
\usepackage{bbm} % double-line symbols (RR), with numerals (11) and lower-case
\usepackage{xspace}                    %%% correct spacing self-defined macros
\usepackage{dcolumn}                    %%% Align table columns on decimal point
\usepackage{bm}                        %%% bold math

\usepackage{slashed}
\ifx\feynVersion\AnswerYes
   \usepackage{feynmp-auto} % for Feynman diagrams; nice definitions after \begin{document}
\fi
\ifx\ShowLabelsVersion\AnswerYes
   \usepackage[color]{showkeys} 
   \definecolor{refkey}{gray}{.5}   % slightly gray font
   \definecolor{labelkey}{gray}{.5} % slightly gray font
   
\fi
\ifx\ShowAnnotationsVersion\AnswerYes
   \newcommand{\comment}[1]{{\scriptsize\sffamily\bfseries{#1}}}
   \newcommand{\margin}[1]{\marginpar{\scriptsize\sffamily\bfseries{#1}}}
   \newcommand{\drafty}{\textbf{Draft version \today} \hfill}
\else
   \newcommand{\comment}[1]{}
   \newcommand{\margin}[1]{}
   \newcommand{\drafty}{}
\fi
\ifx\ShowChangesVersion\AnswerYes
   \usepackage{ulem}
   \newcommand{\delete}[1]{\sout{#1}}            % delete #1 (strike-out)
   \renewcommand{\emph}[1]{\textit{#1}}           % ulem overwrites def of \emph as
\else

   \newcommand{\sout}[1]{}
   \newcommand{\xout}[1]{}

   \newcommand{\delete}[1]{}
\fi

\newcommand{\absatz}{\vspace{2ex}\noindent}

\newcommand{\blue}[1]{#1}
\newcommand{\red}[1]{#1}

\definecolor{green}{rgb}{0.,0.7,0.}
\definecolor{blue}{rgb}{0.,0.,1.0}
\definecolor{red}{rgb}{1,0,0}

\newcommand{\solid}{\rule[0.5ex]{5ex}{1pt}}

\newcommand{\dotdashed}{\rule[0.5ex]{0.6ex}{1pt}\hspace*{0.5ex}$\cdot$\hspace*{0.5ex}\rule[0.5ex]{0.6ex}{1pt}\hspace*{0.5ex}$\cdot$\hspace*{0.5ex}\rule[0.5ex]{0.6ex}{1pt}}

\newcommand{\longdashed}{\rule[0.5ex]{1ex}{1pt}\hspace*{1ex}\rule[0.5ex]{1ex}{1pt}\hspace*{1ex}\rule[0.5ex]{1ex}{1pt}}

\newcommand{\shortdashed}{\rule[0.5ex]{0.5ex}{1pt}\hspace*{0.5ex}\rule[0.5ex]{0.5ex}{1pt}\hspace*{0.5ex}\rule[0.5ex]{0.5ex}{1pt}\hspace*{0.5ex}\rule[0.5ex]{0.5ex}{1pt}\hspace*{0.5ex}\rule[0.5ex]{0.5ex}{1pt}\hspace*{0.5ex}}

\newcommand{\dotted}{$\cdot$\hspace*{0.2ex}$\cdot$\hspace*{0.2ex}$\cdot$\hspace*{0.2ex}$\cdot$\hspace*{0.2ex}$\cdot$\hspace*{0.2ex}$\cdot$\hspace*{0.2ex}}

\newcommand{\dis}{\displaystyle}

\newcommand{\non}{\nonumber}
\newcommand{\hq}{\hspace{0.5em}}

\newcommand{\half}{\frac{1}{2}}
\newcommand{\e}{\mathrm{e}}
\newcommand{\ii}{\mathrm{i}}
\newcommand{\dd}{\mathrm{d}}
\newcommand{\tr}{\mathrm{tr}}

\newcommand{\de}{\partial}

\newcommand{\ev}{\vec{e}}
\newcommand{\kv}{\vec{k}}
\newcommand{\mpi}{\ensuremath{m_\pi}}

\newcommand{\MeV}{\ensuremath{\mathrm{MeV}}}
\newcommand{\fm}{\ensuremath{\mathrm{fm}}}
\newcommand{\ChiEFT}{$\chi$EFT\xspace}

\newcommand{\NXLO}[1]{N\ensuremath{{}^{#1}}LO\xspace}

\newcommand{\HIGS}{HI$\gamma$S\xspace}
\newcommand{\threeHe}{${}^3$He\xspace}

\newcommand{\philin}{\ensuremath{\varphi_\text{lin}}}
\newcommand{\thetad}{\ensuremath{\vartheta_\text{d}}}
\newcommand{\phid}{\ensuremath{\varphi_\text{d}}}

\newcommand{\rhod}{\ensuremath{\rho^{(\dd)}}}
\newcommand{\rhogamma}{\ensuremath{\rho^{(\gamma)}}}

\newcommand{\threej}[6]
    {\ensuremath{\left(\begin{matrix}#1&#2&#3\\#4&#5&#6\end{matrix}\right)}}

\newcommand{\calA}{\mathcal{A}}

\newcommand{\calO}{\mathcal{O}}

\newcommand{\mytitle}[1]{\begin{center}\LARGE{\textbf{#1}}\end{center}}
\newcommand{\myauthor}[1]{\textbf{#1}}
\newcommand{\myaddress}[1]{\textit{#1}}
\newcommand{\mypreprint}[1]{\begin{flushright}#1\end{flushright}}

\newcommand{\alphae}{\ensuremath{\alpha_{E1}}}
\newcommand{\betam}{\ensuremath{\beta_{M1}}}
\newcommand{\gammaee}{\ensuremath{\gamma_{E1E1}}}
\newcommand{\gammamm}{\ensuremath{\gamma_{M1M1}}}
\newcommand{\gammaem}{\ensuremath{\gamma_{E1M2}}}
\newcommand{\gammame}{\ensuremath{\gamma_{M1E2}}}

\begin{document}
%%%%%%%%%%%%%%%%%%%%%%%%%%%%%%%%%%%%%%%%%%%%%%%%%%%%%%%%%%%%%%%%%%%%%%%%%%%%%%%
%%%%%%%%%%%%%%%%%%%%%%%%%%%%%%%%%%%%%%%%%%%%%%%%%%%%%%%%%%%%%%%%%%%%%%%%%%%%%%%
% This is a nice title page including abstract ....
%

\begin{titlepage}
  \setcounter{page}{0} \mypreprint{arXiv:1304.6594 [nucl-th], INT-PUB-13-014\\
    %%%%%%%%%%%%%%%%%%%%%%%%%%%%%%%%%%%%%
    \drafty
    %%%%%%%%%%%%%%%%%%%%%%%%%%%%%%%%%%%%%
    25th April 2013; revised 5th July 2013;
     re-revised with Erratum 26th May 2017\\
     re-re-revised with Second Erratum 11 April 2018\\
%    Re-Revised version 27th May 2005\\
%    Final version 1st July 2005\\
     Eur.~J.~Phys.~\textbf{A49} (2013) 100 with Errata ibid.~\textbf{A53}
     (2017) 113 and \textbf{A54} (2018) 57}
  
\vspace*{-0.5cm}
  
  \mytitle{Dissecting Deuteron Compton Scattering I: The Observables with
    Polarised Initial States}
  
\vspace*{-0.2cm}

\begin{center}
  \myauthor{Harald W.\ Grie\3hammer$^{a,b}$}\footnote{Email:
    hgrie@gwu.edu; permanent address: a}
  
%  \vspace*{0.1cm}
  
  \myaddress{$^a$ Institute for Nuclear Studies, Department of Physics, \\The
    George Washington University, Washington DC 20052, USA} %\\[2ex]

  \emph{and} % \\[2ex]
  \myaddress{$^b$ Institut f\"ur Kernphysik (IKP-3), Institute for Advanced
    Simulation and \\ J\"ulich Centre for Hadron Physics,
    Forschungszentrum J\"ulich, D-52428 J\"ulich, Germany}

%\vspace*{0.2cm}

\end{center}

\vspace*{-0.4cm}

\begin{abstract}
  A complete set of linearly independent observables in Compton scattering
  with arbitrarily polarised real photons off an arbitrarily polarised
  spin-$1$ target is introduced, for the case that the final-state
  polarisations are not measured. Adopted from the one widely used e.g.~in
  deuteron photo-dissociation, it consists of $18$ terms: the unpolarised
  cross section, the beam asymmetry, $4$ target asymmetries and $12$
  asymmetries in which both beam and target are polarised. They are expressed
  by the helicity amplitudes and -- where available -- related to observables
  discussed by other authors. As application to deuteron Compton scattering,
  their dependence on the (isoscalar) scalar and spin dipole polarisabilities
  of the nucleon is explored in Chiral Effective Field Theory with dynamical
  $\Delta(1232)$ degrees of freedom at order $e^2\delta^3$.  Some asymmetries
  are sensitive to only one or two dipole polarisabilities, making them
  particularly attractive for experimental studies.  At a photon energy of
  $100\;\MeV$, a set of $5$ % experimentally realistic but challenging
  observables %, mostly for a tensor-polarised deuteron,
  is identified from which one may be able to extract the spin
  polarisabilities of the nucleon. These are experimentally realistic but
  challenging and mostly involve tensor-polarised deuterons. Relative to
  Compton scattering from a nucleon, sensitivity to the ``mixed'' spin
  polarisabilities $\gammaem$ and $\gammame$ is increased because of
  interference with the $D$ wave component of the deuteron and with its
  pion-exchange current. An interactive \emph{Mathematica 9.0} notebook with
  results for all observables at photon energies up to $120\;\MeV$ is
  available from \texttt{hgrie@gwu.edu}.

  \textbf{Note May 2017/April 2018:} Errors to Eur.~Phys.~J.\ \textbf{A49}
  (2013) 100 were corrected in Errata ibid.~\textbf{A53} (2017) 113 and
  \textbf{A54} (2018) 57 by changing the sign of $T_{11}$ and
  $T_{2(2,1)}^\text{circ}$; interchanging
  $T_{IM}^\text{lin} \to (-)^M\,T_{I,-M}^\text{lin}$; and multiplying
  Eqs.~\eqref{eq:T} and~\eqref{eq:Tcirc} by $(2-\delta_{M0})$. This affects
  Figs.~\ref{fig:screenshot}, \ref{fig:T11varied} to \ref{fig:T21varied},
  \ref{fig:Tcirc11varied}, \ref{fig:Tcirc22varied} to \ref{fig:Tlin11varied},
  \ref{fig:Tlin1-1varied} to \ref{fig:Tlin21varied}, \ref{fig:Tlin2-1varied}
  and \ref{fig:Tlin2-2varied}, and the corresponding, related text. The Errata
  also correct values in Eq.~\eqref{eq:staticpols} to $\gamma_{E1E1}=-5.0$ and
  $\gamma_{M1M1}=3.2$; revert the direction of the vector $\vec{d}$ in
  Fig.~\ref{fig:kinematics}; and cure inconsequential notational
  inconsistencies.

  This arXiv version presents the article corrected for these errors.
\end{abstract}
\vskip-0.1cm
\noindent\small
\begin{tabular}{rl}
  Suggested Keywords: &\begin{minipage}[t]{10.7cm}
    Compton scattering, nucleon polarisabilities, spin
    polarisabilities, deuteron, polarisation observables, complete experiments,
    Chiral Effective Field Theory, $\Delta(1232)$
                    \end{minipage}
\end{tabular}

%\vskip 1.0cm

\end{titlepage}

\setcounter{footnote}{0}

\newpage

%%%%%%%%%%%%%%%%%%%%%%%%%%%%%%%%%%%%%%%%%%%%%%%%%%%%%%%%%%%%%%%%%%%%%%%%%%%%%%%
%%%%%%%%%%%%%%%%%%%%%%%%%%%%%%%%%%%%%%%%%%%%%%%%%%%%%%%%%%%%%%%%%%%%%%%%%%%%%%%
%%%%%%%%%%%%%%%%%%%%%%%%%%%%%%%%%%%%%%%%%%%%%%%%%%%%%%%%%%%%%%%%%%%%%%%%%%%%%%%
% Main Body
%

%%%%%%%%%%%%%%%%%%%%%%%%%%%%%%%%%%%%%%%%%%%%%%%%%%%%%%%%%%%%%%%%%%%%%%%%%%%%%%%
\section{Introduction}
\setcounter{equation}{0}
\label{sec:introduction}

Compton scattering $\gamma \mathrm{X}\to \gamma \mathrm{X}$ at energies below
$1\;\mathrm{GeV}$ explores the two-photon response of the internal low-energy
degrees of freedom in the nucleon and in the lightest nuclei. Since the
electric and magnetic fields of a real photon induce radiation multipoles by
displacing the charged constituents and currents in the target,
energy-dependence and multipolarity of the emitted radiation test the
symmetries and strengths of the interactions between and with them; see a
recent review for details~\cite{Griesshammer:2012we}. In deuteron Compton
scattering, one not only has access to the proton and neutron response, but
also to how photons couple to the charged pion-exchange currents, thus testing
nuclear binding in the simplest stable few-nucleon system. In addition, the
constructive interference with the $D$ wave component of the deuteron can be
expected to lead to increased sensitivity of the hadronic response to the
quadrupole components of the photon fields.

A new generation of high-luminosity facilities like \HIGS, MAMI and MAX-Lab
with near-$100\%$ linear or circular beam polarisation have started to explore
these opportunities. Dense deuteron targets with vector polarisations
approaching $90\%$ are standard.  Since tensor and vector polarisations are
related when in thermal equilibrium with a solid lattice, most
vector-polarised deuteron targets automatically also provide tensor
polarisation degrees of $\lesssim75\%$ -- and the potential for greater values
in dedicated set-ups~\cite{Paetz, crabb, meyer}.

Now is thus an opportune moment for a comprehensive classification of
independent deuteron amplitudes and observables. The spin-$\half$ case has
been discussed by Babusci et al.~\cite{Babusci:1998ww}. For the deuteron,
Chen, Ji and Li~\cite{Chen:2004wwa} constructed a basis for those $12$
amplitudes which remain linearly independent after parity and time-reversal
invariance have been invoked on the $[2$(photon helicities)$\times3$(deuteron
helicities)$]^2$(both in- and out-state)$=36$ helicity amplitudes. However, a
corresponding list of 23 independent observables ($12$ complex amplitudes
minus an overall phase) is missing.  While several single and double
polarisation observables have been constructed and their sensitivity to the
nucleon polarisabilities explored~\cite{Chen:2004wwa, Chen:2004wv,
  Choudhury:2004yz, ShuklaThesis, Griesshammer:2010pz}, no systematic study of
vector and tensor polarisation observables exists. Only one tensor observable
has been considered explicitly, namely for an unpolarised
beam~\cite{Chen:1998rz, Karakowski:1999pt, Karakowski:1999eb}.  What is more,
some deuteron ``vector'' observables which were defined analogous to the
spin-$\half$ case will be shown to actually receive contributions from both
vector and tensor polarisations.

For the case that the polarisations of the final state are not detected, this
work aims to classify all $18$ independent observables and their relation to
the helicity amplitudes. At present, this seems to be the experimentally most
feasible situation. Instead of simply extending the work by Babusci et al.~to
the spin-$1$ case, the starting point is the most general cross section of an
arbitrarily polarised photon beam on an arbitrarily polarised spin-$1$ target,
in a form which is well-known e.g.~from deuteron
photo-disintegration~\cite{Arenhovel:1990yg}. It is parametrised in terms of
the unpolarised cross section, $1$ beam and $4$ target asymmetries as well as
$12$ double asymmetries and has the added benefit that experiments in
less-than-ideal settings can easily be described as well, like when residual
or mixed target and beam polarisations exist. A future publication will define
and study $5$ additional independent polarisation transfer
observables~\cite{hgfuture}. A complete set of independent Compton scattering
observables will then be available from which the $23$ real parameters which
characterise deuteron Compton scattering (i.e.~its independent amplitudes) can
be reconstructed in full.

\absatz The second part of this article explores the sensitivity of the $18$
observables to the two-photon response of the individual nucleon.  Remember
that the proportionality constants between the electric or magnetic field of
the incident photon and the radiation multipoles induced in each nucleon are
the energy-dependent (dynamical) polarisabilities of the
nucleon~\cite{Griesshammer:2001uw, Hildebrandt:2003fm}.  They parametrise the
stiffness of the nucleon $N$ (spin $\frac{\vec{\sigma}}{2}$) against
transitions $\blue{Xl\to Yl^\prime}$ of definite photon multipolarity
% in the electro-magnetic field of a photon with 
at frequency $\omega$ ($l^\prime=l\pm\{0;1\}$; $X,Y=E,M$; $T_{ij}=\half
(\de_iT_j + \de_jT_i)$; $T=E,B$); see e.g.~\cite{Griesshammer:2012we,
  Holstein:2000yj} and references therein. Re-written as point-like
interactions between photons and nucleons, the terms which contain photon
dipoles read:
\begin{equation}
\begin{split}
  \label{eq:polsfromints}
 % \calL_\text{pol}=
  2\pi\;N^\dagger
  \;\big[&\red{\alpha_{E1}(\omega)}\;\vec{E}^2\;+
  \;\red{\beta_{M1}(\omega)}\;\vec{B}^2\; +\;\red{\gamma_{E1E1}(\omega)}
  \;\vec{\sigma}\cdot(\vec{E}\times\dot{\vec{E}})\;
  +\;\red{\gamma_{M1M1}(\omega)}
  \;\vec{\sigma}\cdot(\vec{B}\times\dot{\vec{B}})
  %\non
\\&%\;\;\;\;\;\;\;\;\;\;\;\;
  -\;2\red{\gamma_{M1E2}(\omega)}\;\sigma^i\;B^j\;E_{ij}\;+
  \;2\red{\gamma_{E1M2}(\omega)}\;\sigma^i\;E^j\;B_{ij} \;+\;\dots \big]\;N
\end{split}
\end{equation} 
Since each interaction with a photon leaves a unique signal in such dispersive
effects, Compton scattering allows one to study the symmetries and dynamics of
the hadronic constituents in detail.

The zero-energy values, $\alphae:=\alpha_{E1}(\omega=0)$ etc., are often
quoted as ``the (static) polarisabilities''.  Two scalar %(spin-independent)
polarisabilities $\alpha_{E1}(\omega)$ and $\beta_{M1}(\omega)$ parametrise
electric and magnetic dipole transitions.  The four dipole
spin polarisabilities $\gamma_{E1E1}(\omega)$, $\gamma_{M1M1}(\omega)$,
$\gamma_{E1M2}(\omega)$ and $\gamma_{M1E2}(\omega)$ encode the response of the
nucleon spin-structure. These are particularly interesting since, intuitively
interpreted, they parametrise the bi-refringence which the electromagnetic
field associated with the spin degrees causes in the nucleon, in analogy to
the classical Faraday-effect~\cite{Holstein:2000yj}. The information
accessible in Compton scattering thus goes well beyond that in tests of the
one-photon response e.g.~in form-factor experiments.

Theoretical input is of course needed to carefully evaluate data-consistency
in one model-independent framework for hidden systematic errors; identify the
underlying mechanisms using minimal theoretical bias, like the detailed chiral
dynamics of the pion cloud and of the $\Delta(1232)$ as the lowest nucleon
resonance; and, most importantly, explain how these findings emerge from QCD
by relating to emerging lattice simulations (see most recently~\cite{De10,
  Engelhardt:2011qq, Lujan:2011ue}).  The polarisabilities also enter as one
of the bigger sources of uncertainties in theoretical determinations of the
proton-neutron mass shift (see e.g.~most recently \cite{WalkerLoud:2012bg})
and of the two-photon-exchange contribution to the Lamb shift in muonic
hydrogen \cite{Pachucki, Carlson:2011dz, Bi12}. While presumably not providing
a solution to the proton-charge-radius puzzle, they also contribute in
radiative corrections to this process, see e.g.~\cite{Miller:2012ne}. For all
these goals, Chiral Effective Field Theory ($\chi$EFT), the low-energy theory
of QCD and extension of Chiral Perturbation Theory to few-nucleon systems,
adds objective estimates of the theoretical uncertainties. Indeed, \ChiEFT has
been particularly successful in describing proton and few-nucleon Compton
scattering, starting with the first calculation and sensitivity study of the
scalar polarisabilities in \ChiEFT~\cite{Bernard:1991rq, Bernard:1995dp}.
Ref.~\cite{Griesshammer:2012we} contains details on its history and status in
Compton scattering, as well as on \ChiEFT variants not discussed here.

Having established a consistent database from all available proton and
deuteron data below $350\;\MeV$ in Ref.~\cite{Griesshammer:2012we}, the static
scalar polarisabilities of the proton were recently extracted in this framework
with a $\chi^2$ per degree of freedom of $113/135$~\cite{McGovern:2012ew}:
\begin{equation}
  \label{eq:p1parameterfinalfit}
  \begin{split}
    \alphae^{(\text{p})} &=10.7\pm
    0.3(\text{stat})\pm0.2(\text{Baldin})\pm0.3(\text{theory})
    \\
    \betam^{(\text{p})} &=\phantom{0}3.1\mp
    0.3(\text{stat})\pm0.2(\text{Baldin})\mp0.3(\text{theory})
  \end{split}
\end{equation}
Throughout, polarisabilities without superscripts denote isoscalar quantities,
and the canonical units of $10^{-4}\;\fm^3$ for scalar and $10^{-4}\;\fm^4$
for spin dipole polarisabilities are understood.

Since the deuteron is an isoscalar, elastic scattering on it provides of
course only access to the isoscalar (average) nucleon polarisabilities.  In
Ref.~\cite{Griesshammer:2012we}, these were found to have much larger errors
since deuteron data is less accurate and more scarce (with
$\chi^2/$d.o.f.$=24/25$):
\begin{equation}
  \label{eq:d1parameterfinalfit}
  \begin{split}
    \alphae &=10.9\pm
    0.9(\text{stat})\pm0.2(\text{Baldin})\pm0.8(\text{theory})
    \\
    \betam &=\phantom{0}3.6\mp
    0.9(\text{stat})\pm0.2(\text{Baldin})\mp0.8(\text{theory})
  \end{split}
\end{equation} 
These results were derived using the Baldin sum rules, whose isoscalar variant
reads~\cite{Griesshammer:2012we}:
\begin{equation}
  \label{eq:isoscalarBaldin} \alphae+\betam=14.5\pm0.3\;\;.
\end{equation}
These publications also discuss in detail the fit procedure and residual
theoretical uncertainties. Comparing Eqs.~\eqref{eq:p1parameterfinalfit}
and~\eqref{eq:d1parameterfinalfit} shows that within the data-dominated error,
the two-photon responses of the proton and neutron as parametrised by the
scalar polarisabilities are identical. A particularly interesting prediction
of \ChiEFT is that small proton-neutron differences stem from chiral-symmetry
breaking interactions with and in the pion cloud around the nucleon, probing
details of QCD.  Experiments are therefore underway and planned to improve the
Compton scattering database; see e.g.~\cite{Griesshammer:2012we} for details.
Their other focus are the spin polarisabilities. Only the linear combinations
$\gamma_0$ and $\gamma_\pi$ of scattering under $0^\circ$ and $180^\circ$ are
somewhat constrained by data or phenomenology.  Conflicting results from MAMI
and LEGS exist for the proton, and large error-bars are found for the
neutron~\cite{Schumacher:2005an}. The isoscalar values are in the range (see
also~\cite{Griesshammer:2010pz}):
\begin{equation}
\begin{split}
\label{eq:gamma0pi}
  \gamma_{0}&:=-{\gamma}_{E1E1}-{\gamma}_{M1M1}-
  {\gamma}_{E1M2}-{\gamma}_{M1E2}\approx0\\
  \gamma_{\pi}&:=-{\gamma}_{E1E1}+{\gamma}_{M1M1}-{\gamma}_{E1M2}+
  {\gamma}_{M1E2}\approx[5\dots15]
\end{split}
\end{equation}
A comprehensive classification of independent amplitudes and observables is
thus warranted, including a detailed study of dependencies on scalar and spin
polarisabilities. Insofar, this publication extends the so-far most thorough
work in Ref.~\cite{Griesshammer:2010pz}, including its Erratum.

\absatz After defining the most general cross section without detection of the
polarisations of the final state in Subsec.~\ref{sec:parameterising}, the
remainder of Sec.~\ref{sec:construction} is devoted to the more technical
issues of relating its observables to the helicity amplitudes of deuteron
Compton scattering and to other parameter-combinations found in the
literature, including the Babusci-classification.
Section~\ref{sec:observables} discusses the sensitivity of the observables to
the dipole polarisabilities, with an eye towards potential experiments.  It
also proposes a road-map to the isoscalar, spin-independent and spin-dependent
nucleon polarisabilities from high-accuracy experiments with deuteron targets.
A customary summary in Sec.~\ref{sec:conclusions} rounds off the article.
Preliminary results were presented in a recent
proceeding~\cite{Griesshammer:2013eha}.

%%%%%%%%%%%%%%%%%%%%%%%%%%%%%%%%%%%%%%%%%%%%%%%%%%%%%%%%%%%%%%%%%%%%%%%%%%%%%%%
\section{Constructing Observables}
\setcounter{equation}{0}
\label{sec:construction}

\subsection{Kinematics and Polarisation States}
\label{sec:kinematics}

\begin{figure}[!htb]
\begin{center}
  \includegraphics[width=0.55\textwidth]{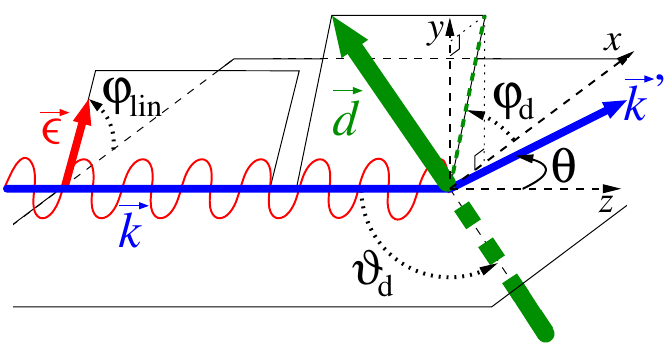}
  \caption{(Colour on-line) Kinematics of deuteron Compton scattering:
    incoming photon along the $z$-axis, linearly polarised at angle
    $\varphi_\text{lin}$ relative to the scattering plane ($xz$-plane);
    scattering angle $\theta$; deuteron polarisation axis $\vec{d}$ with
    azimuthal angle $\vartheta_\text{d}$ from the $z$-axis to $\vec{d}$, and
    polar angle $\varphi_\text{d}$ from the $x$-axis to the projection of
    $\vec{d}$ onto the $xy$-plane; $y$-axis the normal of the scattering
    plane; $\kv$ ($\kv^\prime$) the momentum of the incident (outgoing)
    photon.}
\label{fig:kinematics}
\end{center}
\end{figure}

This presentation follows the reviews of Arenh\"ovel and
Sanzone~\cite{Arenhovel:1990yg}, and Paetz~\cite{Paetz}. Inspired by the
former, the kinematics is pictorially represented in
Fig.~\ref{fig:kinematics}.
The photon beam polarisation is described by a density matrix with entries
\begin{equation}
  \label{eq:photonpol}
  \left(\rhogamma\right)_{\lambda^\prime\lambda}:=
  \langle \lambda^\prime|\rhogamma|\lambda\rangle
  =\frac{1}{2}\left[
    \delta_{\lambda^\prime\lambda}\left(1+\lambda\;P^{(\gamma)}_\text{circ}
    \right)
    -\delta_{\lambda^\prime,-\lambda}\;P^{(\gamma)}_\text{lin}\;
    \e^{+2\lambda\ii\philin}\right]\;\;.
\end{equation}
Here, $P^{(\gamma)}_\text{circ}\in[-1;1]$ is the degree of right-circular
polarisation, i.e.~the difference between right and left circular
polarisation, with $P^{(\gamma)}_\text{circ}=+1/-1$ describing a fully
right/left circularly polarised photon (positive/negative helicities
$\lambda,\lambda^\prime=\pm$ by
$\ev_\pm=-\frac{\ii}{\sqrt{2}}(\ev_y\mp\ii\ev_x)$). The degree of linear
polarisation is parametrised by $P^{(\gamma)}_\text{lin}\in[0;1]$, with
$\philin\in[0;\pi[$ the angle from the $x$-axis to the polarisation
plane\footnote{This definition varies from that of~\cite{Arenhovel:1990yg},
  whose angle $\phi$ is counted \emph{from} the polarisation plane \emph{to}
  the normal of the scattering plane, i.e.~$\philin=-\phi$.}, i.e.~a photon
polarisation $\vec{\epsilon}_\text{lin}=\ev_x\cos\philin+\ev_y\sin\philin$.

Today's deuteron targets are both vector- and tensor-polarised along the same
axis~\cite{Paetz}. Let the axis $\vec{d}$ in which $\rhod$ is diagonal be
oriented as in Fig.~\ref{fig:kinematics}, i.e.\footnote{This definition varies
  from that of~\cite{Arenhovel:1990yg}, whose angles are defined as
  $\phi_\text{d}-\phi=\phid$, but still $\theta_\text{d}=\thetad$.}
\begin{equation}
\label{eq:dpol}
  \vec{d}=\left(\begin{array}{c}\sin\thetad\;\cos\phid\\
                               \sin\thetad\;\sin\phid\\
                               \cos\thetad\end{array}\right)
\end{equation} 
with azimuthal angle $\thetad\in[0;\pi]$ and polar angle $\phid\in[0;2\pi]$.
The entries of the polarisation density matrix are then in the basis
$M_{\vec{d}}=(1;0;-1)$ of magnetic quantum numbers along $\vec{d}$:
\begin{equation}
  \label{eq:deuteronpol1}
  \rhod_{\vec{d}}=\frac{1}{3}\left[
    P^{(\dd)}_0\;1\;
    +\sqrt{\frac{3}{2}}\;P^{(\dd)}_1
    \left(\begin{array}{ccc}
      1&0&0\\0&0&0\\0&0&-1
    \end{array}\right)
    +\frac{1}{\sqrt{2}}\;P^{(\dd)}_2
    \left(\begin{array}{ccc}
      1&0&0\\0&-2&0\\0&0&1
    \end{array}\right)
\right]\;\;.
\end{equation}
The subscript denotes of course that the system is quantised along the
$\vec{d}$-axis, not the $z$-axis, and is kept for comparison with the
literature~\cite{Paetz, Arenhovel:1990yg}.  Here, $P^{(\dd)}_0:=1$
parametrises the part of the deuteron density matrix which behaves like a
scalar under rotations, while $P^{(\dd)}_1$ and $P^{(\dd)}_2$ characterise the
parts which transform like an (irreducible) spherical vector and tensor
operator, respectively. These can be related to the degrees of vector and
tensor polarisation in Cartesian coordinates: along this quantisation axis, a
fraction $1\ge p_{\pm,0}\ge0$ populates the state with magnetic quantum number
$M_{\vec{d}}=\pm1,0$. The overall norm is $p_++p_-+p_0=1$. The degree of
vector polarisation is in Cartesian coordinates
$P_z=p_+-p_-=\sqrt{\frac{2}{3}}\;P^{(\dd)}_1\in[-1;1]$, and that of tensor
polarisation is $P_{zz}=p_++p_--2p_0=1-3p_0=\sqrt{2}\;P^{(\dd)}_2\in[1;-2]$.
Because $p_{\pm,0}$ lies between 0 and 1, they are subject to the combined
constraint $2\sqrt{2}\ge P^{(\dd)}_2+\sqrt{3}\,|P^{(\dd)}_1|\ge-\sqrt{2}$.
When the deuteron spins are in thermal equilibrium with a solid lattice,
tensor and vector polarisations are related by $P_{zz}=2-\sqrt{4-3P_z^2}$,
i.e.~$P^{(\dd)}_2=\sqrt{2}-\sqrt{2-(P^{(\dd)}_1)^2}$~\cite{Paetz, crabb}.

The advantage to decompose $\rhod$ into irreducible representations of the
rotation group is that it is then particularly simple to change the
quantisation axis from $\vec{d}$ to the beam axis
$\hat{\kv}:=\vec{k}/\omega\equiv\ev_z$; cf.~\cite[Subsec.~13]{Rose}.
Ref.~\cite{Arenhovel:1990yg} finally provides the angular momentum
representation of the spin-$1$ polarisation density matrix which is diagonal
along $\vec{d}$:
\begin{equation}
  \label{eq:deuteronpol}
  \rhod_{m^\prime m}:=\langle m^\prime|\rhod|m\rangle=
  \frac{(-1)^{1-m}}{\sqrt{3}}\sum\limits_{I=0}^{2}\sqrt{2I+1}\;P^{(\dd)}_I
    \sum\limits_{M=-I}^I \threej{1}{1}{I}{m}{-m^\prime}{-M}\e^{\ii M\phid}\;
    d^I_{M0}(\thetad)\;\;.
\end{equation}
The conventions for $3j$-symbols and reduced Wigner-$d$ matrices are those of
Rose~\cite{Rose} and Edmonds~\cite{Edmonds}, also listed in the Particle Data
Booklet~\cite{PDG}.

%%%%%%%%%%%%%%%%%%%%%%%
\subsection{Parametrising the Cross Section}
\label{sec:parameterising}

Like any reaction $\gamma\mathrm{d}\to\mathrm{X}$, deuteron Compton scattering
and deuteron photo-disintegration share the same in-state. As long as the
final-state polarisations are not detected (i.e.~are summed over), their
differential cross sections are thus characterised by the same dependence on
the initial-state deuteron and photon polarisations. One can therefore adopt
the decomposition familiar from deuteron
photo-disintegration~\cite{Arenhovel:1990yg} to Compton scattering:
\begin{equation}
\begin{split}
  \label{eq:arencross}
  \frac{\dd\sigma}{\dd\Omega}=\left.\frac{\dd\sigma}{\dd\Omega}\right|_\text{unpol}\;\bigg[&1\;
  +\;\Sigma^\text{lin}(\omega,\theta)\;P^{(\gamma)}_\text{lin}\;\cos2\philin\\
  &+ \sum\limits_{I=1,2\atop 0\le M\le I}\;T_{IM}(\omega,\theta)\;
  P^{(\dd)}_I\;d^I_{M0}(\thetad)\;\cos[M\phid-\frac{\pi}{2}\delta_{I1}]\\
  &+ \sum\limits_{I=1,2\atop 0\le M\le I}\;T_{IM}^\text{circ}(\omega,\theta)\;
  P^{(\dd)}_I\;d^I_{M0}(\thetad)\;
  P^{(\gamma)}_\text{circ}\;\sin[M\phid+\frac{\pi}{2}\delta_{I1}]\\
  &+ \sum\limits_{I=1,2\atop -I\le M\le I}\;T_{IM}^\text{lin}(\omega,\theta)\;
  P^{(\dd)}_I\;d^I_{M0}(\thetad)\;
  P^{(\gamma)}_\text{lin}\;\cos[M\phid-2\philin-\frac{\pi}{2}\delta_{I1}]
  \bigg]
\end{split}
\end{equation}
Besides the trivial limitations $I\in\{0;1;2\}$ and $|M|\leq I$, the
summations in Eq.~\eqref{eq:arencross} are easily shown to be constrained by
trivial zeros and double counting of angular dependencies:
\begin{itemize}

\item $T_{00}\equiv1$, i.e.~the first factor in Eq.~\eqref{eq:arencross} could
  also be written as $1\equiv T_{00}\,P^{(\dd)}_0$;

\item $T_{IM}=(-)^{I+M}\,T_{I,-M}$, and in particular $T_{10}\equiv0$; 

\item $T^\text{circ}_{IM}=(-)^{I+M+1}\,T^\text{circ}_{I,-M}$, and in
  particular $T^\text{circ}_{00}\equiv0$ (the circular-beam asymmetry on an
  unpolarised target, identical zero due to rotation invariance) and
  $T^\text{circ}_{20}\equiv0$;

\item $T^\text{lin}_{00}\equiv\Sigma^\text{lin}$.
\end{itemize}
The cross section is thus fully parametrised by the following linearly
independent functions:
\begin{itemize}

\item $1$ differential cross section
  $\dis\left.\frac{\dd\sigma}{\dd\Omega}\right|_\text{unpol}$ of unpolarised
  photons on an unpolarised target;

\item $1$ beam asymmetry of a linearly polarised beam on an unpolarised
  target $\Sigma^\text{lin}$;

\item $1$ vector target asymmetry %(recoil polarisation)
  on an unpolarised beam $T_{11}$;

\item $3$ tensor target asymmetries %(recoil polarisations)
  of an unpolarised beam $T_{2M}$, $M=0,1,2$;

\item $2$ double asymmetries of circular photons on a vector polarised
  target $T^\text{circ}_{1M}$, $M=0,1$;

\item $2$ double asymmetries of circular photons on a tensor polarised
  target $T^\text{circ}_{2M}$, $M=1,2$;

\item $3$ double asymmetries of linear photons on a vector target
  $T^\text{lin}_{1M}$, $M=0,\pm1$;

\item $5$ double asymmetries of linear photons on a tensor target
  $T^\text{lin}_{2M}$, $M=0,\pm1,\pm2$.
\end{itemize}
Since these 18 real, independent functions of scattering energy and angle are
of course process-dependent, those discussed in Compton scattering differ from
those in e.g.~deuteron photo-disintegration. The decomposition of
Eq.~\eqref{eq:arencross} holds in any frame, but the functions are
frame-dependent. It also applies when the polarisation of the target and/or
scattered photon is detected in the final state, without specifying the
initial state. The 18 recoil polarisations are thus identical to the functions
above.

%%%%%%%%%%%%%%%%%%%%%%%
\subsection{Matching Helicity Amplitudes to Observables}
\label{sec:matching}

Deuteron Compton scattering amplitudes $T$ are usually described in the
helicity basis (dependencies on $\omega,\theta$ and other parameters are
dropped for brevity in this Section):
\begin{equation}
  \label{eq:basis}
  A^{M_f\lambda_f}_{M_i\lambda_i}:=\langle M_f,\lambda_f|T|
  M_i,\lambda_i \rangle\;\;, 
\end{equation}
where $\lambda_{i/f}=\pm$ is the circular polarisation of the initial/final
photon, and $M_{i/f}\in\{0;\pm1\}$ is the magnetic quantum number of the
initial/final deuteron spin. In the following, the indices and summations over
the final-state polarisations are suppressed as self-understood,
e.g.~$A_{M_i\lambda_i}\equiv A^{M_f\lambda_f}_{M_i\lambda_i}$. In addition, it
is convenient to introduce an abbreviation for the sum over all polarisations
of the squared amplitude:
\begin{equation}
  \label{eq:cala}
  |\calA|^2:=\sum\limits_{M_i,\lambda_i} |A_{M_i\lambda_i}|^2\equiv
  \sum\limits_{M_f,\lambda_f; M_i,\lambda_i}
  |A^{M_f\lambda_f}_{M_i\lambda_i}|^2\;\;.
\end{equation} 
The cross section of Compton scattering of a photon beam with the density
matrix $\rhogamma$ from a target with density matrix $\rhod$, without
detection of the final state polarisations, is then
\begin{equation} 
\label{eq:xsect}
  \frac{\dd\sigma}{\dd\Omega}=\Phi^2\;\tr[T\rhod\rhogamma T^\dagger]\;\;,
\end{equation}
where the trace is taken over the polarisation states and $\Phi$ is the
frame-dependent flux factor, e.g.~in the centre-of-mass and lab frames:
\begin{equation}
  \Phi_{\mathrm{cm}} = \frac{M_\mathrm{d}}{4 \pi}\;
    \frac{1}{\omega_\text{cm}+\sqrt{M_\text{d}^2+\omega_\text{cm}^2}} \quad ,
    \quad \Phi_{\mathrm{lab}} = \frac{M_\mathrm{d}}{4\pi}
    \;\frac{1}{M_\mathrm{d}+\omega_\text{lab}(1-\cos\theta_\text{lab})}\;\;.
\end{equation}
Transformations between lab and cm kinematics are found in a recent
review~\cite[Sec.~2.3]{Griesshammer:2012we}.

By inserting the density matrices of Eqs.~\eqref{eq:photonpol}
and~\eqref{eq:deuteronpol} into Eq.~\eqref{eq:xsect}, one obtains the cross
section in terms of the amplitudes, as function of photon polarisations
$P^{(\gamma)}_\text{circ}$ and $P^{(\gamma)}_\text{lin}$ with polarisation
angle $\philin$ and deuteron polarisation $P^{(\dd)}_I$ with orientation
$(\thetad,\phid)$. The functional dependence of the result on these parameters
is easily matched to the parametrisation in Eq.~\eqref{eq:arencross}. For the
unpolarised part, one finds of course:
\begin{equation}
  \label{eq:cross}
  \left.\frac{\dd\sigma}{\dd\Omega}\right|_\text{unpol}=\frac{\Phi^2}{6}
  |\calA|^2\;\;. 
\end{equation}
The asymmetries are then (these definitions obey the constraints discussed in
Sec.~\ref{sec:parameterising}): 
\begin{eqnarray}
  \label{eq:sigmalin}
%p.66
  \Sigma^\text{lin}\;|\calA|^2\!\!\!&=&\!\!\!-\sum\limits_{M_i,\lambda_i}
  A_{M_i\lambda_i}A^*_{M_i,-\lambda_i}\\
%  p.66
  \label{eq:T}
  T_{IM}\;|\calA|^2\!\!\!&=&\!\!\!\sqrt{3(2I+1)}\;\ii^{\delta_{I1}}%\ii^{I-1}
  \;(2-\delta_{M0})\!\!\!\!
  \sum\limits_{M_i,M_i^\prime,\lambda_i}\!\!\!\!  (-)^{1-M_i}\!%\;
  \threej{1}{1}{I}{M_i}{-M_i^\prime}{-M}%\;
  \!\!A_{M_i^\prime\lambda_i}A^*_{M_i\lambda_i}\\
% p.67
  \label{eq:Tcirc}
  T_{IM}^\text{circ}\;|\calA|^2\!\!\!&=&\!\!\!\sqrt{3(2I+1)}\;%\sqrt{3}\;\sqrt{2I+1}\;
  \ii^{\delta_{I2}}%\ii^{1-\delta_{I1}}%\ii^{I-1}
  (2-\delta_{M0})\!\!\!\!\!
  \sum\limits_{M_i,M_i^\prime,\lambda_i}\!\!\!  (-)^{1-M_i}\lambda_i%\;
  \!\threej{1}{1}{I}{M_i}{-M_i^\prime}{-M}\!\!%\;
  A_{M_i^\prime\lambda_i}A^*_{M_i\lambda_i}\\
% p.65
  \label{eq:Tlin}
  T_{IM}^\text{lin}\;|\calA|^2\!\!\!&=&\!\!\!\sqrt{3(2I+1)}%\sqrt{3}\;\sqrt{2I+1}
  \!\!\!\sum\limits_{M_i,M_i^\prime,\lambda_i}\!\!\!
  (-)^{-M_i}
 \left(\ii\lambda_i\right)^{\delta_{I1}}\!\!\lambda_i^M
  \threej{1}{1}{I}{M_i}{-M_i^\prime}{-\lambda_iM}
  A_{M_i^\prime\lambda_i}A^*_{M_i,-\lambda_i}
\end{eqnarray}
These explicit forms can also be used to determine which observables are
nonzero only due to inelasticities.  Cross sections and, concurrently, the
functions $\Sigma^\text{lin}$, $T_{IM}$, $T_{IM}^\text{circ/lin}$ are of
course real.  The Compton amplitudes $A_{M_i\lambda_i}$ are real below the
first inelasticity, so that the occurrence of the imaginary unit in six of the
observables in Eqs.~\eqref{eq:sigmalin} to~\eqref{eq:Tlin} indicates that they
are zero there, namely
\begin{equation}
\label{eq:zerobelow}
\mbox{below the first inelasticity: }
T_{11}\equiv0\;\;,\;\; T_{2(1,2)}^\text{circ}\equiv0\;\;,\;\; 
T_{1(0,\pm1)}^\text{lin}\equiv0.
\end{equation}

%%%%%%%%%%%%%%%%%%%%%%%
\subsection{Complete Experiments?}
\label{sec:complete}

% As discussed in the Introduction, t
The deuteron Compton amplitude contains 2 independent complex amplitudes for a
scalar target, 4 more for a vector target, and 6 more for a tensor target; see
e.g.~\cite{Chen:2004wwa}. How many and which of them are accessible with
polarised beam and/or target, but without measuring outgoing polarisations
(or, by time-reversal invariance, vice versa)? Those which cannot be
determined must be probed in polarisation transfer experiments. These are
significantly harder because of the difficulties to measure recoil and
scattered-photon polarisations.

As a warm-up, one could consider first the Compton scattering below the first
inelasticity, where all amplitudes are real. This is however of limited use
in deuteron Compton scattering, where the first appreciable breakup process,
$\gamma\mathrm{d}\to\mathrm{p}\mathrm{n}$, starts at a cm photon energy of
$B_\text{d}=2.225\;\MeV$, namely so low that the amplitudes have significant
imaginary parts in the experimentally interesting region\footnote{The first
  inelasticity opens at zero energy, with multiple photons in the final state
  ($\gamma\mathrm{d}\to\gamma\gamma\mathrm{d}$ etc.), but is suppressed by
  powers of $\alpha=1/137$ and hence does not significantly contribute in
  experiments. It is not considered in today's theoretical descriptions, whose
  first inelasticity thus is the deuteron breakup.}. In contradistinction, the
first appreciable inelasticity on the proton starts at the one-pion production
threshold.

Above the first inelasticity, $23$ independent real amplitudes exist, namely 3
for a scalar target (2 complex minus an overall phase), 8 more for a vector
target, and 12 more for a tensor target. Since the 6 observables of
Eq.~\eqref{eq:zerobelow} are nonzero there, one finds:
\begin{itemize}
\item For scalar targets, only 2 of 3 observables are accessible, leaving
  1 to be determined from a polarisation transfer observable. 

\item For vector polarised targets, 6 of 8 observables are accessible, leaving
  2 to be determined from polarisation transfer observables.

\item For tensor polarised targets, 10 of 12 observables are accessible,
  leaving again 2 to be determined from polarisation transfer observables.
\end{itemize}
The $5$ correlations between beam and recoiling target polarisation which are
necessary for complete experiments on the deuteron will be discussed in a
future publication~\cite{hgfuture}.

This concludes the classification itself; results in \ChiEFT will be presented
in Sec.~\ref{sec:results}.

%%%%%%%%%%%%%%%%%%%%%%%
\subsection{Relation to Other Parametrisations}
\label{sec:context}

Since some observables in Compton scattering with vector and tensor polarised
targets have been constructed before, it is appropriate to relate these to the
classification in Eq.~\eqref{eq:arencross}. Often, observables are expressed
not in terms of the degrees of deuteron vector and tensor polarisations, but
via the occupation numbers $p_{\pm,0}$ of a state quantised along $\vec{d}$.
From Eq.~\eqref{eq:dpol}, the density matrix of a pure deuteron state
$|M_{\vec{d}}\rangle$ is:
\begin{equation}
\label{eq:MJtoP}
  \begin{array}{r@{:\hq}llll}
    p_\pm=1
    &\rhod_{\vec{d}}=|M_{\vec{d}}=\pm1\;\rangle\langle\;M_{\vec{d}}=\pm1\;|
    %\delta_{M_{f\vec{d}}M_{i\vec{d}}}\delta_{M_{i\vec{d}},\pm1} 
      &\Longleftrightarrow&
      \dis P^{(\dd)}_1=\pm\sqrt{\frac{3}{2}}&\mbox{and }\dis P^{(\dd)}_2=
      \frac{1}{\sqrt{2}}\\ 
     p_0=1  &\rhod_{\vec{d}}=|M_{\vec{d}}=0\;\rangle\langle\;M_{\vec{d}}=0\;|
      %\delta_{M_{f\vec{d}} M_{i\vec{d}}}\delta_{M_{i\vec{d}} 0}
        &\Longleftrightarrow& 
      \dis P^{(\dd)}_1=0&\mbox{and }\dis P^{(\dd)}_2= -\sqrt{2}\;\;.
 \end{array}
\end{equation}

\subsubsection{Chen's Tensor-Polarised Cross Section~\cite{Chen:1998rz}}
\label{sec:chen}

The first tensor observable was constructed by Chen~\cite{Chen:1998rz}, and
also used by Karakowski and Miller~\cite{Karakowski:1999pt,Karakowski:1999eb}.
His definition of a cross section combination for an unpolarised beam on a
deuteron which is tensor polarised along the $z$ axis translates into the
observables of Eq.~\eqref{eq:arencross} with
$P^{(\gamma)}_\text{circ}=P^{(\gamma)}_\text{lin}=0$, $\thetad=\phid=0$ and
Eq.~\eqref{eq:MJtoP} into:
\begin{equation} 
\label{eq:dsigma2chen}
  \frac{\dd\sigma_2^\text{\cite{Chen:1998rz}}}{\dd\Omega}:=
  \frac{1}{4}\left[  
    2\frac{\dd\sigma}{\dd\Omega}(M_{iz}=0)-
    \frac{\dd\sigma}{\dd\Omega}(M_{iz}=1)
    -\frac{\dd\sigma}{\dd\Omega}(M_{iz}=-1)\right]=-\frac{3}{2\sqrt{2}}\;T_{20}
  \left.\frac{\dd\sigma}{\dd\Omega}\right|_\text{unpol}\;\;,
\end{equation}
where the subscript in $M_{iz}$ denotes that $\vec{d}$ points along the
$z$-axis for the initial state. From now on, the bracketed superscript of an
observable indicates the bibliographic reference from which the notation is
taken verbatim.

This is the only tensor-observable for which calculations exist, namely at
$49$ and $69\;\MeV$ both by Chen and by Karakowski and Miller. Nonetheless,
these will not be compared in detail with those of the \ChiEFT approach taken
in Sec.~\ref{sec:results}. Chen's are derived in ``pion-less'' EFT, i.e.~for
typical momenta well below the pion mass and typical photon energies
$\omega\lesssim\mpi^2/M\approx20\;\MeV$~\cite{Griesshammer:2012we}.  These
predictions are thus more of qualitative interest. Shape and size of the
angular dependence differ indeed considerably from those presented later.
Karakowski and Miller used an approach similar to that which will be outlined
in Sec.~\ref{sec:theory}, but without a dynamical $\Delta(1232)$ and without
some pion-exchange diagrams dictated by chiral
symmetry~\cite{Karakowski:1999pt,Karakowski:1999eb}. Their results at $49$ and
$69\;\MeV$ agree up to about $30\%$ in shape and magnitude with the ones
presented below. The difference does not stem from the $\Delta(1232)$, but may
be attributed to the fact that their photon-nucleon interaction for
rescattering terms is expanded only to first order, while Hildebrandt et
al.~demonstrated that terms up to $l=2$ should be kept for
convergence~\cite{Hildebrandt:2005iw,Hildebrandt:2005ix}. Tensor-observables
should be more susceptible to this difference.

\subsubsection{Scalar and Vector Target Observables by Babusci et
  al.~\cite{Babusci:1998ww}}

Babusci et al.~\cite{Babusci:1998ww} were the first to identify a complete set
of independent observables for Compton scattering, namely for a spin-$\half$
target. Their classification applies of course also to a scalar- or
vector-polarised deuteron target, % $P^{(\dd)}_1=0$ or $1$,
\emph{provided} one sets the tensor-component to zero, $P^{(\dd)}_2\equiv0$.
Following the discussion in Sec.~\ref{sec:kinematics}, this constrains
$|P^{(\text{d})}_1|\le\sqrt{\frac{2}{3}}$, so the vector polarisation cannot
reach the maximal value of $1$ allowed for a spin-$\frac{1}{2}$ target. While
one should be aware of this difference, we choose in the following to quote
results with a pretense value ``$P^{(\text{d})}_1=1$''. Those for a deuteron
target which is maximally vector polarised but not tensor polarised are
obtained from these by multiplying the right-hand sides of
Eqs.~\eqref{eq:sigmaybab} to \eqref{eq:sigma3ybab} by $\sqrt{\frac{2}{3}}$.

Experimentally, these observables are measured as asymmetries between cross
sections with different target and beam polarisation angles
$(\thetad,\phid;\philin)$, %and $(\thetad,\phid;\philin)$,
normalised to their sum. The configurations are chosen such that their cross
sections sum to twice the total unpolarised cross section.

Specifically, the beam asymmetry in Refs.~\cite{Babusci:1998ww,
  Choudhury:2004yz, ShuklaThesis, Griesshammer:2010pz} is the difference of
the cross sections of a linearly polarised beam ($P^{(\gamma)}_\text{lin}=1$,
$P^{(\gamma)}_\text{circ}=0$) either in the scattering plane ($\philin=0$) or
perpendicular to it ($\philin=\pi/2$) on an unpolarised target
($P^{(\dd)}_1=P^{(\dd)}_2=0$), normalised to their sum.  Inserting these
choices into Eq.~\eqref{eq:arencross} identifies
\begin{eqnarray}
\label{eq:sigma3bab}
  \Sigma_3^\text{\cite{Babusci:1998ww}}&\equiv& \Sigma^\text{\cite{Choudhury:2004yz, ShuklaThesis}}
  \equiv\Pi^\text{lin~\cite{Griesshammer:2010pz}}=
  \frac{\dis \frac{\dd\sigma}{\dd\Omega}(\philin=0)-
    \frac{\dd\sigma}{\dd\Omega}(\philin=
    \frac{\pi}{2})}{\dis \frac{\dd\sigma}{\dd\Omega}(\philin=0)+
    \frac{\dd\sigma}{\dd\Omega}(\philin=\frac{\pi}{2})}\non\\
  &\equiv&
  \frac{(\philin=0)-(\philin=\frac{\pi}{2})}{(\philin=0)+(\philin=\frac{\pi}{2})}
  \equiv
  \frac{(\philin=0)-(\philin=\frac{\pi}{2})}{.\;+\;.}\\
  &=&
  \Sigma^\text{lin}\;\;.\non
\end{eqnarray}
For readability, the differential cross section symbol is dropped in each term
in the second line, and an abbreviation ``$.\;+\;.$'' is introduced for a
denominator which is the sum, rather than the difference, of the terms in the
numerator. Not surprisingly, all definitions of the beam asymmetry
coincide.

The vector target asymmetry with unpolarised beam ($P^{(\dd)}_1=1$,
$P^{(\dd)}_2=P^{(\gamma)}_\text{circ}=P^{(\gamma)}_\text{lin}=0$)
translates as
\begin{equation}
  \label{eq:sigmaybab}
  \Sigma_y^\text{\cite{Babusci:1998ww}}=
  \frac{(\thetad=\frac{\pi}{2},\phid=+\frac{\pi}{2})-
    (\thetad=\frac{\pi}{2},\phid=-\frac{\pi}{2})}{.\;+\;.}
  =-\frac{1}{\sqrt{2}}\;T_{11}\;\;,
\end{equation}
the vector target asymmetries with right-circularly polarised beam
($P^{(\dd)}_1=1$, $P^{(\gamma)}_\text{circ}=1$,
$P^{(\dd)}_2=P^{(\gamma)}_\text{lin}=0$) as
\begin{eqnarray}
\label{eq:sigma2xbab}
  \Sigma_{2x}^\text{\cite{Babusci:1998ww}}&=&
  \frac{(\thetad=\frac{\pi}{2},\phid=0)-
              (\thetad=\frac{\pi}{2},\phid=\pi)}{.\;+\;.} 
  =-\frac{1}{\sqrt{2}}\;T_{11}^\text{circ}\\
\label{eq:sigma2zbab}
  \Sigma_{2z}^\text{\cite{Babusci:1998ww}}&=&
  \frac{(\thetad=0)-
              (\thetad=\pi)}{.\;+\;.} 
  = T_{10}^\text{circ}\;\;,
\end{eqnarray}
and finally those with linearly polarised beam on a vector target
($P^{(\dd)}_1=1$, $P^{(\gamma)}_\text{lin}=1$,
$P^{(\dd)}_2=P^{(\gamma)}_\text{circ}=0$) as
\begin{eqnarray}
\label{eq:sigma1xbab}
  \Sigma_{1x}^\text{\cite{Babusci:1998ww}}&=&
  \frac{(\thetad=\frac{\pi}{2},\phid=0;\philin=+\frac{\pi}{4})-
      (\frac{\pi}{2},0;\philin=-\frac{\pi}{4})}{.\;+\;.}
  =\frac{1}{\sqrt{2}}\left(T_{11}^\text{lin}-T_{1,-1}^\text{lin}\right)\\
\label{eq:sigma1zbab}
  \Sigma_{1z}^\text{\cite{Babusci:1998ww}}&=&
  \frac{(\thetad=0;\philin=+\frac{\pi}{4})-
      (0;\philin=-\frac{\pi}{4})}{.\;+\;.}   
  =-T_{10}^\text{lin}\\
\label{eq:sigma3ybab}
  \Sigma_{3y}^\text{\cite{Babusci:1998ww}}&=&
  \frac{[(\thetad=\frac{\pi}{2},\phid=\frac{\pi}{2};\philin=0)-
    (\frac{\pi}{2},\frac{\pi}{2};\frac{\pi}{2})]-
 [(\frac{\pi}{2},-\frac{\pi}{2};0)-
 (\frac{\pi}{2},-\frac{\pi}{2};\frac{\pi}{2})]}
{[\;.\;+\;.\;]\;+\;[\;.\;+\;.\;]}  
  \non\\&
  =&-\frac{1}{\sqrt{2}}\;
  \left(T_{11}^\text{lin}+T_{1,-1}^\text{lin}\right)\;\;,
\end{eqnarray}
or 
\begin{equation}
  T_{11}^\text{lin}=\frac{1}{\sqrt{2}}\left(
    \Sigma_{1x}^\text{\cite{Babusci:1998ww}}-\Sigma_{3y}^\text{\cite{Babusci:1998ww}}\right)
  \;\;,\;\;
  T_{1,-1}^\text{lin}=-\frac{1}{\sqrt{2}}\left(
    \Sigma_{1x}^\text{\cite{Babusci:1998ww}}+\Sigma_{3y}^\text{\cite{Babusci:1998ww}}\right)\;\;.
\end{equation}
In $\Sigma_{2x/z}$, Babusci et al.~flip the circular beam polarisation. Due
to parity symmetry, this is equivalent to the flip of the target polarisation
performed above. 

\subsubsection{Polarised Deuteron Observables by Chen et
  al.~\cite{Chen:2004wwa}, Choudhury/Phillips\ \cite{Choudhury:2004yz,
    ShuklaThesis} and Grie\3hammer/Shukla~\cite{Griesshammer:2010pz}}
\label{sec:shuklaobs}

These authors define observables in analogy to those introduced by Babusci et
al.~\cite{Babusci:1998ww}. However, the deuteron is taken to be prepared such
that only the magnetic quantum numbers $M_{i\vec{d}}=\pm1$ contribute, in the
direction $\vec{d}$ in which the density matrix is diagonal. To understand why
this difference may lead to confusion, consider the single-polarisation
observable for scattering an unpolarised (or circularly polarised) beam on a
deuteron target which is polarised in a pure $M_{iy}=\pm1$ state perpendicular
to the scattering plane (i.e.~parallel or anti-parallel to the $y$ axis):
\begin{equation}
  \label{eq:sigmaydef}
  \Sigma_y^\text{\cite{Chen:2004wwa}}=
  \frac{\dis\frac{\dd\sigma}{\dd\Omega}(M_{iy}=+1)-
    \frac{\dd\sigma}{\dd\Omega}(M_{iy}=-1)}{.\;+\;.}\;\;,
\end{equation}
where the same abbreviation as in Eq.~\eqref{eq:sigma3bab} is used.  This
appears to be the natural application of
$\Sigma_{y}^\text{\cite{Babusci:1998ww}}$, Eq.~\eqref{eq:sigmaybab}, to the
deuteron.  Since the deuteron polarisation is flipped in the difference, the
numerator should describe a vector-polarised deuteron.  According to
Eq.~\eqref{eq:MJtoP}, a pure state $|M_{i\vec{d}}|=1$ is described by
$P^{(\dd)}_1=\sqrt{3/2}$ and $P^{(\dd)}_2=1/\sqrt{2}$. For this observable,
$\vec{d}$ is parallel to the $y$-axis, so that $M_{iy}=\pm1$ corresponds to
$\thetad=\pi/2,\phid=\pm\pi/2$.  With
$P^{(\gamma)}_\text{circ}=P^{(\gamma)}_\text{lin}=0$ and the same
abbreviations as before, the numerator becomes:
\begin{equation}
  (M_{iy}=+1)-(M_{iy}=-1)=
  (\thetad=\frac{\pi}{2},\phid=+\frac{\pi}{2})-
  (\thetad=\frac{\pi}{2},\phid=-\frac{\pi}{2})=
  -\sqrt{3} \;T_{11}
  \left.\frac{\dd\sigma}{\dd\Omega}\right|_\text{unpol}.
\end{equation}
Tensor-observables do indeed not contribute. In contradistinction, the
denominator reads:
\begin{eqnarray}
  (M_{iy}=+1)+(M_{iy}=-1)&=&(\thetad=\frac{\pi}{2},\phid=+\frac{\pi}{2})+
  (\thetad=\frac{\pi}{2},\phid=-\frac{\pi}{2})\non\\
  &=&
  \left[2-\left(\frac{1}{\sqrt{2}}\;T_{20}+\frac{\sqrt{3}}{2}\;T_{22}\right)
  \right]\left.
    \frac{\dd\sigma}{\dd\Omega}\right|_\text{unpol}\;\;.
\end{eqnarray}
It is no more proportional to the unpolarised cross section since the
$M_{iy}=0$-term is absent, as noted already in
Refs.~\cite{Chen:2004wwa,Choudhury:2004yz, ShuklaThesis}. Like in
$\Sigma_{y}^\text{\cite{Babusci:1998ww}}$ of Eq.~\eqref{eq:sigmaybab}, the
resulting asymmetry
\begin{equation}
\label{eq:sigmaywrong}
  \Sigma_y^\text{\cite{Chen:2004wwa}}=-\frac{2\sqrt{3}\;T_{11}}
  {4-\sqrt{3}\;T_{22}-\sqrt{2}\;T_{20}}
\end{equation}
is proportional to $T_{11}$, but the prefactor has changed and now depends in
addition on the tensor-polarised observables $T_{2(0,2)}$. While the same
symbol is used for the vector target polarisation in
Ref.~\cite{Babusci:1998ww} and for that of Ref.~\cite{Chen:2004wwa},
Eq.~\eqref{eq:sigmaydef}, the two are actually different:
\begin{equation}
  \Sigma_y^\text{\cite{Chen:2004wwa}}\neq\Sigma_y^\text{\cite{Babusci:1998ww}}\;\;!
\end{equation}
It is for that reason that the apparent notational degeneracy is lifted
throughout this article by including an explicit reference
superscript. 

Translating the other observables of Refs.~\cite{Chen:2004wwa,
  Choudhury:2004yz, ShuklaThesis, Griesshammer:2010pz} is now straightforward.
Asymmetries with unpolarised targets are of course identical,
Eq.~\eqref{eq:sigma3bab}. Since Ref.~\cite{Griesshammer:2010pz} considers both
differences of polarised cross sections (denoted by
$\Delta^\text{\cite{Griesshammer:2010pz}}$) and their asymmetries
$\Sigma^\text{\cite{Griesshammer:2010pz}}$, both are also recorded in the
following. One finds with $P^{(\dd)}_1=\sqrt{3/2}$, $P^{(\dd)}_2=1/\sqrt{2}$,
$P^{(\gamma)}_\text{circ}=1$ and $P^{(\gamma)}_\text{lin}=0$ for the
asymmetries built in analogy to $\Sigma_{2x/z}^\text{\cite{Babusci:1998ww}}$:
\begin{eqnarray}
\label{eq:deltaxcircwrong}
  \Delta_x^\text{circ~\cite{Griesshammer:2010pz}}&=&
  (M_{ix}=+1;\lambda_i=1)-(M_{ix}=-1;1)=
  (\thetad=\frac{\pi}{2},\phid=0)-(\frac{\pi}{2},\phid=\pi)\non\\
  &=&-\sqrt{3}\;T_{11}^\text{circ}
  \left.\frac{\dd\sigma}{\dd\Omega}\right|_\text{unpol}\\
\label{eq:sigmaxcircwrong}
  \Sigma_x^\text{circ~\cite{Griesshammer:2010pz}}&\equiv&\Sigma_x^\text{\cite{Chen:2004wwa,Choudhury:2004yz, ShuklaThesis}}=
  \frac{\Delta_x^\text{circ~\cite{Griesshammer:2010pz}}}{.\;+\;.}
  =-\frac{2\sqrt{3}\;T_{11}^\text{circ}}
  {4+\sqrt{3}\;T_{22}-\sqrt{2}\;T_{20}}\\
\label{eq:deltazcircwrong}
  \Delta_z^\text{circ~\cite{Griesshammer:2010pz}}&\equiv&
  2[\Delta_1\frac{\dd\sigma}{\dd\Omega}]^\text{\cite{Chen:2004wwa}}=
  (M_{iz}=+1;1)-(M_{iz}=-1;1)=(\thetad=0)-(\thetad=\pi)\non\\
  &=&\sqrt{6} \;T_{10}^\text{circ}
  \left.\frac{\dd\sigma}{\dd\Omega}\right|_\text{unpol}\\
\label{eq:sigmazcircwrong}
\Sigma^\text{circ~\cite{Griesshammer:2010pz}}_z&\equiv&\Sigma_z^\text{\cite{Choudhury:2004yz, ShuklaThesis}}\equiv-\Sigma_z^\text{\cite{Chen:2004wwa}}=
\frac{\Delta_z^\text{circ~\cite{Griesshammer:2010pz}}}{.\;+\;.}=
\frac{\sqrt{3}\;T_{10}^\text{circ}}{\sqrt{2}+T_{20}}
\end{eqnarray}
In no case is the denominator just proportional to the unpolarised cross
section; instead, it also depends on $T_{2(0,\pm2)}$. It should be noted that
Ref.~\cite{Chen:2004wwa} provides formulae for the denominators of
$\Sigma_{x/y/z}^\text{\cite{Chen:2004wwa}}$ which depend only on the scalar
and vector parts of the target polarisation. These results could not be
reproduced.

The following additional cross section differences and asymmetries for
linearly polarised beam on a polarised deuteron target were described in
Ref.~\cite{Griesshammer:2010pz}:
\begin{eqnarray}
\label{eq:deltaxlinwrong}
  \Delta^\text{lin~\cite{Griesshammer:2010pz}}_x&=&
  (M_{ix}=1;\philin=0)-(M_{ix}=1;\philin=\frac{\pi}{2})\non\\
  &=&
  (\thetad=\frac{\pi}{2},\phid=0;\philin=0)-(\frac{\pi}{2},0;\frac{\pi}{2})
  \non\\
  &=&\left[2\;\Sigma^\text{lin}+\frac{\sqrt{3}}{2}
    \left(T_{22}^\text{lin}+T_{2,-2}^\text{lin}\right)
    -\frac{1}{\sqrt{2}}\;T_{20}^\text{lin}
  \right]\left.\frac{\dd\sigma}{\dd\Omega}\right|_\text{unpol}\\
\label{eq:sigmaxlinwrong}
  \Sigma_x^\text{lin~\cite{Griesshammer:2010pz}}&=&
  \frac{\Delta_x^\text{lin~\cite{Griesshammer:2010pz}}}{.\;+\;.}
  =\frac{4\;\Sigma^\text{lin}+
    \sqrt{3}\left(T_{22}^\text{lin}+T_{2,-2}^\text{lin}\right)
    -\sqrt{2}\;T_{20}^\text{lin}}
  {4+\sqrt{3}\;T_{22}-\sqrt{2}\;T_{20}}\\
\label{eq:deltazlinwrong}
  \Delta^\text{lin~\cite{Griesshammer:2010pz}}_z&=&
  (M_{iz}=1;\philin=0)-(M_{iz}=1;\philin=\frac{\pi}{2})=
  (\thetad=0;\philin=0)-(0;\frac{\pi}{2})\non\\
  &=&\left[2\;\Sigma^\text{lin}+\sqrt{2}\;T^\text{lin}_{20}\right]
  \left.\frac{\dd\sigma}{\dd\Omega}\right|_\text{unpol}\\
\label{eq:sigmazlinwrong}
  \Sigma_z^\text{lin~\cite{Griesshammer:2010pz}}&=&
  \frac{\Delta_z^\text{lin~\cite{Griesshammer:2010pz}}}{.\;+\;.}
  =\frac{2\;\Sigma^\text{lin}+\sqrt{2}\;T^\text{lin}_{20}}{2+\sqrt{2}\;T_{20}}
\end{eqnarray}
with $P^{(\dd)}_1=\sqrt{3/2}$, $P^{(\dd)}_2=1/\sqrt{2}$,
$P^{(\gamma)}_\text{circ}=0$ and $P^{(\gamma)}_\text{lin}=1$. Notice that the
numerators $\Delta_{x/z}^\text{lin~\cite{Griesshammer:2010pz}}$ depend on
different and nontrivial combinations of both $\Sigma^\text{lin}$ and
$T_{2(0,\pm2)}^\text{lin}$.  $\Sigma_x^\text{lin~\cite{Griesshammer:2010pz}}$
and $\Sigma_z^\text{lin~\cite{Griesshammer:2010pz}}$ would be identical if the
tensor-polarised observables were zero.

The additional terms proportional to $T_{2(0,2)}$ in each denominator of
$\Sigma_y^\text{\cite{Chen:2004wwa}}$ and
$\Sigma^\text{circ/lin~\cite{Griesshammer:2010pz}}_{x/z}$ will by themselves
turn out to be rather large, sensitive to the polarisabilities, and
significantly dependent on photon energy and scattering angle; see
Figs.~\ref{fig:compareTs}, \ref{fig:T22varied} and \ref{fig:T20varied} in
Sec.~\ref{sec:results}. Without this input, no simple conclusions can thus be
drawn how the sensitivity of $\Sigma_y^\text{\cite{Chen:2004wwa}}$ on the
polarisabilities translates into the sensitivity of its numerator alone.  On
the other hand, $\Delta^\text{lin~\cite{Griesshammer:2010pz}}_{x/z}$ is
dominated by $\Sigma^\text{lin}$ and $T^\text{lin}_{22}$ since
$T^\text{lin}_{2(0,-2)}$ will turn out to be very small.

%%%%%%%%%%%%%%%%%%%%%%%%%%%%%%%%%%%%%%%%%%%%%%%%%%%%%%%%%%%%%%%%%%%%%%%%%%%%%%%
\section{Observables in \ChiEFT}
\setcounter{equation}{0}
\label{sec:observables}

%%%%%%%%%%%%%%%%%%%%%%%%%%%%%%%%%%%%%%%%%%%%%%%%%%%%%%%%%%%%%%%%%%%%%%%%%%%%%%%
\subsection{Theoretical Ingredients}
\label{sec:theory}

The following sub-sections explore the sensitivity of the $18$ independent
observables to the scalar and spin dipole polarisabilities in \ChiEFT. Since
this version of the deuteron Compton scattering amplitudes is described
comprehensively in previous publications~\cite{Hildebrandt:2005iw,
  Hildebrandt:2005ix, Griesshammer:2010pz} and summarised in a recent
review~\cite{Griesshammer:2012we}, its main ingredients are only sketched
here.

In \ChiEFT with explicit $\Delta(1232)$ degrees of freedom, four typical
low-energy scales are found in deuteron Compton scattering: the pion mass
$\mpi\approx 140\;\MeV$ as the typical chiral scale; the Delta-nucleon mass
splitting $\Delta_M\approx290\;\MeV$; the deuteron binding momentum (inverse
deuteron size) $\gamma\approx45\;\MeV$ as the typical scale of the bound
$\mathrm{NN}$ system; and the photon energy $\omega$. When measured in units
of a natural ``high'' scale $\Lambda_\chi\gg\Delta_M,\mpi,\omega,\gamma$ at
which \ChiEFT with explicit $\Delta(1232)$ degrees of freedom can be expected
to break down because new degrees of freedom become dynamical, each gives rise
to a small, dimensionless expansion parameter.  Typical values of
$\Lambda_\chi$ are the masses of the $\omega$ and $\rho$ as the next-lightest
exchange mesons (about $700\;\MeV$). To avoid a fourfold expansion, it is
convenient to approximately identify some scales so that only one
dimensionless parameter is left. In the $\delta$-expansion of Pascalutsa and
Phillips~\cite{Pascalutsa:2002pi}, one chooses
\begin{equation}
  \delta\equiv
  \frac{\Delta_M}{\Lambda_\chi}\approx\left(\frac{m_\pi}{\Lambda_\chi}
  \right)^{1/2}\qquad,
  \label{eq:delta}
\end{equation}
i.e.~numerically $\delta\approx0.4$. The identity is exact for
$\Lambda_\chi\approx 600\;\MeV$. Since present experiments are run at
$\omega\lesssim200\;\MeV$, % one sets $\omega\lesssim\mpi$ and
the nonzero Delta-width is not tested, cf.~Ref.~\cite{McGovern:2012ew}.

The two-nucleon dynamics adds the momentum scale $\gamma$ of the shallow bound
state. Based on Refs.~\cite{Griesshammer:2004yn,Griesshammer:2010pz,menu07,
  Hildebrandt:2005iw,Hildebrandt:2005ix}, Chapter 5 of
Ref.~\cite{Griesshammer:2012we} provides a ``unified'' deuteron Compton
amplitude which is complete at order $e^2\delta^3$ and valid from zero photon
energy to just below the pion production threshold, $\omega\lesssim\mpi$. This
variant is identical to $\calO(\epsilon^3)$ in the ``Small Scale
Expansion''~\cite{Jenkins:1990jv, Jenkins:1991ne,
  Hemmert:1996xg,Hemmert:1997ye}, used in Ref.~\cite{Griesshammer:2010pz}. At
this order, the Compton scattering kernel consists of ``one-nucleon
contributions'' in which both photons interact with the same nucleon
(Fig.~\ref{fig:onenucleon}), and ``two-nucleon contributions''
(Fig.~\ref{fig:twonucleon}). The latter consists of two classes, each of which
contributes at the order $\calO(e^2\delta^3)$ of the present formulation:
\begin{figure}[!htb]
\begin{center}
      \includegraphics[width=0.82\textwidth]{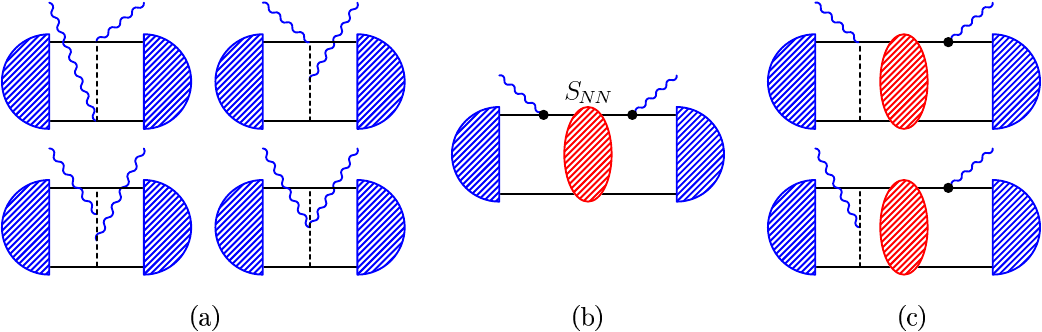}
      \caption{(Colour on-line) Two-nucleon contributions in \ChiEFT up to
        order $e^2\delta^3$ (permuted and crossed diagrams not shown). Photons
        couple to the same pion (a); rescattering contributions (b,c).
        Ellipse: two-nucleon $S$-matrix; dot: coupling via minimal
        substitution or magnetic moment; from
        Ref.~\cite{Griesshammer:2010pz}.}
    \label{fig:twonucleon}
\end{center}
\end{figure}
\begin{enumerate}
\item Both photons couple to the charged pion-exchange current,
  Fig.~\ref{fig:twonucleon} (a)~\cite{Beane:1999uq}.
\item Each photon couples to the nucleon charge, magnetic moment and/or to
  different pion-exchange currents; Fig.~\ref{fig:twonucleon} (b) and (c).
  Between the two couplings, the nucleons rescatter arbitrarily often via the
  full NN S-matrix (including no rescattering at all). These contributions are
  small for $\omega\sim\mpi$ but required for $\omega\lesssim\gamma$ in order
  to restore the exact low-energy theorem of Compton scattering, i.e.~the
  Thomson limit~\cite{Friar:1975, Friar:1977zq, Arenhovel:1980jx,
    Weyrauch:1984tf}.  At zero energy, its emergence in the \ChiEFT
  power-counting mandates that the contribution of Fig.~\ref{fig:twonucleon}
  (b) must be exactly minus half that of the one-nucleon Thomson term,
  Fig.~\ref{fig:onenucleon} (a), and that the pion-exchange contributions of
  Fig.~\ref{fig:twonucleon} (a) and (c) must add to
  zero~\cite{Griesshammer:2012we}. Such stringent numerical tests are
  fulfilled to better than $0.2\%$. At higher energies, the significance of
  this cancellation belies in a considerable reduction of the dependence of
  the amplitudes on the deuteron wave function and $\mathrm{NN}$
  potential~\cite{Griesshammer:2012we}.
\end{enumerate}

The one-nucleon sector is formed by:
\begin{figure}[!htb]
\begin{center}
     \includegraphics[width=0.82\textwidth]{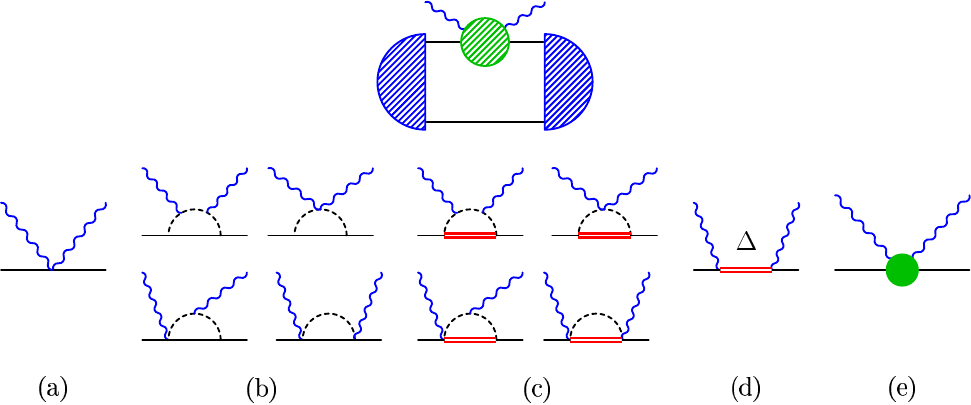}
     \caption{(Colour on-line) One-nucleon contributions in \ChiEFT up to
       $\calO(e^2\delta^3)$ (permuted and crossed diagrams not shown). Top:
       embedding into the deuteron. Bottom: one-nucleon Thomson term (a); pion
       cloud of the nucleon (b) and $\Delta(1232)$ (double line; (c));
       excitation of an intermediate $\Delta$ (d); short-distance effects to
       $\alphae$ and $\betam$ (e); from Ref.~\cite{Griesshammer:2010pz}.}
    \label{fig:onenucleon}
\end{center}
\end{figure}
\begin{enumerate}
\item Single-nucleon Thomson scattering, Fig.~\ref{fig:onenucleon} (a), is the
  leading-order term, $\calO(e^2\delta^0)$.
\item Coupling to the chiral dynamics of the single-nucleon pion cloud,
  Fig.~\ref{fig:onenucleon} (b),  $\calO(e^2\delta^2)$.
\item Excitation of the $\Delta(1232)$ intermediate state,
  Fig.~\ref{fig:onenucleon} (d), and coupling to the pion cloud around it,
  Fig.~\ref{fig:onenucleon} (c), each contributing at $\calO(e^2\delta^3)$ for
  $\omega\lesssim\mpi$. Following Ref.~\cite{Griesshammer:2012we}, the
  $\Delta$ is treated non-relativistically and with zero width, using
  $\Delta_M=293\;\MeV$, $g_{\pi\mathrm{N}\Delta}=1.425$ and the
  non-relativistic version of the $\mathrm{N}\Delta\gamma$ $M1$-coupling
  $b_1=5$, obtained from converting the relativistic value of $g_M=2.9$. This
  value, in turn, is found by fitting the single-nucleon amplitudes to the
  data above $150\;\MeV$ in the proton Compton database established there.
\item Two energy-independent, isoscalar short-distance coefficients,
  Fig.~\ref{fig:onenucleon} (e), which encode those contributions to the
  nucleon polarisabilities $\alphae$ and $\betam$ which arise at this order
  neither from pions nor from the $\Delta(1232)$. Since they are formally of
  one order higher, $\mathcal{O}(e^2{\delta}^4)$, the order of the resulting
  total amplitude is called ``modified
  $\calO(e^2\delta^3)$''.  While these ``off-sets''
  for the static polarisabilities are determined by data, the energy- and
  isospin-dependence of the spin-independent polarisabilities are at this
  order predicted in \ChiEFT. Here, their values are taken from the
  determination in Ref.~\cite{Griesshammer:2012we}; see
  Eq.~\eqref{eq:d1parameterfinalfit}.
\end{enumerate}
Nucleon polarisabilities arise solely from terms (2) to (4). In this
power-counting, ``switching off'' $\Delta(1232)$ contributions is equivalent
to a calculation at one lower order, $\calO(e^2\delta^2)$, in which the scalar
polarisabilities are parameter-free predictions:
$\alphae=10\betam=12.5$~\cite{Bernard:1991rq}.

These kernels are convoluted with deuteron wave functions to obtain the
amplitudes $%A^{M_f\lambda_f}_{M_i\lambda_i}\equiv
\langle M_f,\lambda_f|T| M_i,\lambda_i \rangle$ of Eq.~\eqref{eq:basis}.
Results in this article are obtained with the \ChiEFT deuteron wave function
at \NXLO{2} (cutoff $650\;\MeV$) in the implementation of Epelbaum et
al.~\cite{Epelbaum:1999dj} and the AV18 potential~\cite{av18} for NN
rescattering. This combination provides an adequate \ChiEFT representation of
the two-nucleon system; see discussion in Ref.~\cite{Griesshammer:2012we} and
Sec.~\ref{sec:moredependences} below.

This formulation differs from the previous ones of
Refs.~\cite{Hildebrandt:2005ix,Hildebrandt:2005iw, Griesshammer:2010pz} in
some numerical improvements, a new parameter set $(b_1,\,g_{\pi
  \mathrm{N}\Delta},\,\Delta_M)$ for the $\Delta(1232)$ from the Breit-Wigner
parameters and the proton Compton data, and in slightly changed numbers for
the isoscalar, scalar polarisabilities. In a fully consistent EFT calculation,
the kernel, wave functions and potential should of course be derived in the
same framework. This is work in progress.

%%%%%%%%%%%%%%%%%%%%%%%%%%%%%%%%%%%%%%%%%%%%%%%%%%%%%%%%%%%%%%%%%%%%%%%%%%%%%%%
\subsection{Strategy}
\label{sec:strategy}

At this (modified) order $e^2\delta^3$, the static isoscalar dipole
polarisabilities are (with theoretical uncertainties of about $\pm0.8$ from
higher-order contributions and in the canonical units of $10^{-4}\;\fm^3$ for
the scalar polarisabilities and $10^{-4}\;\fm^4$ for the spin-dependent
ones)~\cite{Hildebrandt:2003fm,Griesshammer:2010pz,Griesshammer:2012we}:
\begin{equation}
  \begin{split}
  \label{eq:staticpols}
  {\alpha}_{E1} = 10.9\;\;&, \;\;{\beta}_{M1} = 3.6  \\
  \gamma_{E1E1} = -5.0\;\;, \;\;
  \gamma_{M1M1} = 3.2 \;\;&, \;\;
  \gamma_{M1E2} = 0.9 \;\;, \;\;
  \gamma_{E1M2} = 0.9\;\;,
  \end{split}
\end{equation}
i.e.~$\gamma_0=+0.4$, $\gamma_\pi=8.4$, which is not incompatible with those
of other approaches, see Eq.~\eqref{eq:gamma0pi}.  The values for the spin
polarisabilities differ slightly from these quoted in
Refs.~\cite{Hildebrandt:2003fm,Griesshammer:2010pz} because of the updates to
the $\mathcal{O}(e^2\delta^3)$ amplitudes described in Sect.~5.3 of
Ref.~\cite{Griesshammer:2012we}.

The convergence of the
spin polarisabilities from $\calO(e^2\delta^2)$ via the $\calO(e^2\delta^3)$
values quoted above to the values at $\calO(e^2\delta^4)$ is complicated; see
Table 4.2 of Ref.~\cite{Griesshammer:2012we}. Therefore, no theoretical
uncertainty is assigned for now.

Since the deuteron is an isoscalar, only average nucleon polarisabilities are
accessible in elastic deuteron Compton scattering. In order to analyse the
sensitivity of each observable, one varies each dipole polarisability about
the static central value by adding the parameters $\delta \alpha_{E1}$,
$\delta \beta_{M1}$, $\delta \gamma_{E1E1}$, $\delta \gamma_{M1M1}$, $\delta
\gamma_{E1M2}$ and $\delta \gamma_{M1E2}$ to the interactions of the
single-nucleon sub-system, Eq.~\eqref{eq:polsfromints}~\cite{Choudhury:2004yz,
  Griesshammer:2010pz}.  Their contribution to the amplitudes in the
$\gamma$N cm system is
\begin{eqnarray}
  {A}^{\mathrm{fit}}(\omega,\,z)&=& 4\pi\,\omega^2\,\bigg[
% A1
  [\delta\alpha_{E1}+z\,\delta\beta_{M1}]
  \,(\vec{\epsilon}^\prime\cdot\vec{\epsilon})
% A2
  -\delta\beta_{M1}\,
  (\vec{\epsilon}^\prime\cdot\hat{k})\,(\vec{\epsilon}\cdot\hat{k}^\prime)\nonumber\\
% A3
  &&-\ii\,[\delta\gamma_{E1E1}+z\,\delta\gamma_{M1M1}+\delta\gamma_{E1M2}
         +z\,\delta\gamma_{M1E2}]\,\omega\,\vec{\sigma}\cdot
         (\vec{\epsilon}^\prime\times\vec{\epsilon})\nonumber\\
% A4
  &&+\ii\,[\delta\gamma_{M1E2}-\delta\gamma_{M1M1}]\,\omega\,\vec{\sigma}\cdot
         \left(\hat{k}^\prime\times\hat{k}\right)
         (\vec{\epsilon}^\prime\cdot\vec{\epsilon})\nonumber\\
% A5
  &&+\ii\,\delta\gamma_{M1M1}\,\omega\,\vec{\sigma}\cdot
  \left[\left(\vec{\epsilon}^\prime\times\hat{k}\right)
    (\vec{\epsilon}\cdot\hat{k}^\prime)-
    \left(\vec{\epsilon}\times\hat{k}^\prime\right)
    (\vec{\epsilon}^\prime\cdot\hat{k})\right]\nonumber\\
% A6
  &&+\ii\,\delta\gamma_{E1M2}\,\omega\,\vec{\sigma}\cdot
  \left[\left(\vec{\epsilon}^\prime\times\hat{k}^\prime\right)
    (\vec{\epsilon}\cdot\hat{k}^\prime)-
    \left(\vec{\epsilon}\times\hat{k}\right)
    (\vec{\epsilon}^\prime\cdot\hat{k})\right]
\bigg]\;\;.
\label{eq:fit}
\end{eqnarray}
These variables may be considered as parametrising the difference between
predicted and (so-far un-measured) experimental static values of the
polarisabilities, under the assumption that the energy-dependence from the
pion-cloud and $\Delta(1232)$ is correctly predicted in \ChiEFT.
Alternatively, one can view them as parametrising deviations from the
order-$\e^2\delta^3$ \ChiEFT amplitudes at fixed nonzero energy, including the
theoretical uncertainties of higher-order effects. In that case, the
deviations themselves could be seen as energy-dependent. Such an approach
forms the basis of a multipole analysis of deuteron Compton scattering
advocated in
Refs.~\cite{Griesshammer:2004yn,Miskimentalk,Griesshammer:2010pz}.
Determining the six dipole polarisabilities is then in principle reduced to a
multipole-analysis of $6+1$ high-accuracy scattering experiments.

The variation of the isoscalar values by $\pm2$ canonical units is chosen
since it is roughly at the level of the combined statistical, theoretical and
Baldin-sum-rule induced error for $\alphae$ and
$\betam$~\eqref{eq:d1parameterfinalfit}. With quadratic contributions of the
polarisabilities $\delta(\alphae,\betam,\gamma_i)$ suppressed in the squared
amplitudes, variations by other amounts are easily linearly extrapolated. In
practise, the scalar polarisabilities of the proton are constrained to better
than $\pm2$, so that deuteron Compton scattering experiments are more likely
focused on extracting neutron polarisabilities. In that case, these studies
can be interpreted as providing the sensitivities on varying the neutron
polarisabilities by $\pm4$ units, with fixed proton values.

The spin polarisabilities are however less well known; besides the constraints
of Eq.~\eqref{eq:gamma0pi}, no experimental information
has been published thus far, and theoretical descriptions easily disagree by
as much as $2$ units~\cite{Griesshammer:2012we}. For example, a recent
determination of the scalar dipole polarisabilities of the proton included
varying one of the spin polarisabilities to $\gammamm=2.2\pm0.5(\text{stat})$,
which -- combined with its theoretical accuracy -- would by itself already
suggest a variation by about $2$ units.

Amplitudes from scalar polarisabilities scale like $\omega^2$, while those
containing spin polarisabilities scale like $\omega^3$; see
Eq.~\eqref{eq:fit}.  Ideally, one can therefore perform high-accuracy
experiments at relatively low energies, $\omega\lesssim70\;\MeV$, to better
determine $\alphae$ and $\betam$ and constrain high-energy predictions. The
spin polarisabilities are then extracted at $\gtrsim100\;\MeV$, as already
advocated in Ref.~\cite{Griesshammer:2010pz}. The observables considered here
follow this pattern.

Additionally, one should address:
\begin{enumerate}
\item The Baldin sum rule constraint, Eq.~\eqref{eq:isoscalarBaldin}. However,
  its independent test by better data at forward angles would be expedient.
\item Weaker constraints for the forward and backward spin polarisabilities,
  Eq.~\eqref{eq:gamma0pi}. These come with considerable theoretical and
  systematic uncertainties.
\item Logistic constraints like detector placement and available beam
  energies, as well as detector and polarisation efficiencies. All these must
  be taken into account to determine which experiments have the potential for
  the greatest sensitivity on a given polarisability and of the greatest
  impact in the network of data already available.
\end{enumerate}

Considering asymmetries removes many systematic experimental uncertainties,
but the corresponding count rates are necessary for beam-time estimates and
follow from multiplying with the unpolarised cross section,
cf.~\eqref{eq:arencross}. In general, asymmetries are by $\lesssim30$\% less
sensitive to variations of the polarisabilities than the corresponding count
rates.  Sometimes, sensitivity to the nucleon structure is even lost entirely,
while an enhancement appears in no case. It is the purview of our experimental
colleagues to determine to what extent such draw-backs outweigh the benefits of
measuring asymmetries instead of cross-section differences.
  
\begin{figure}[!htb]%[!htb]
  \begin{center}
    \includegraphics*[width=\linewidth]
       {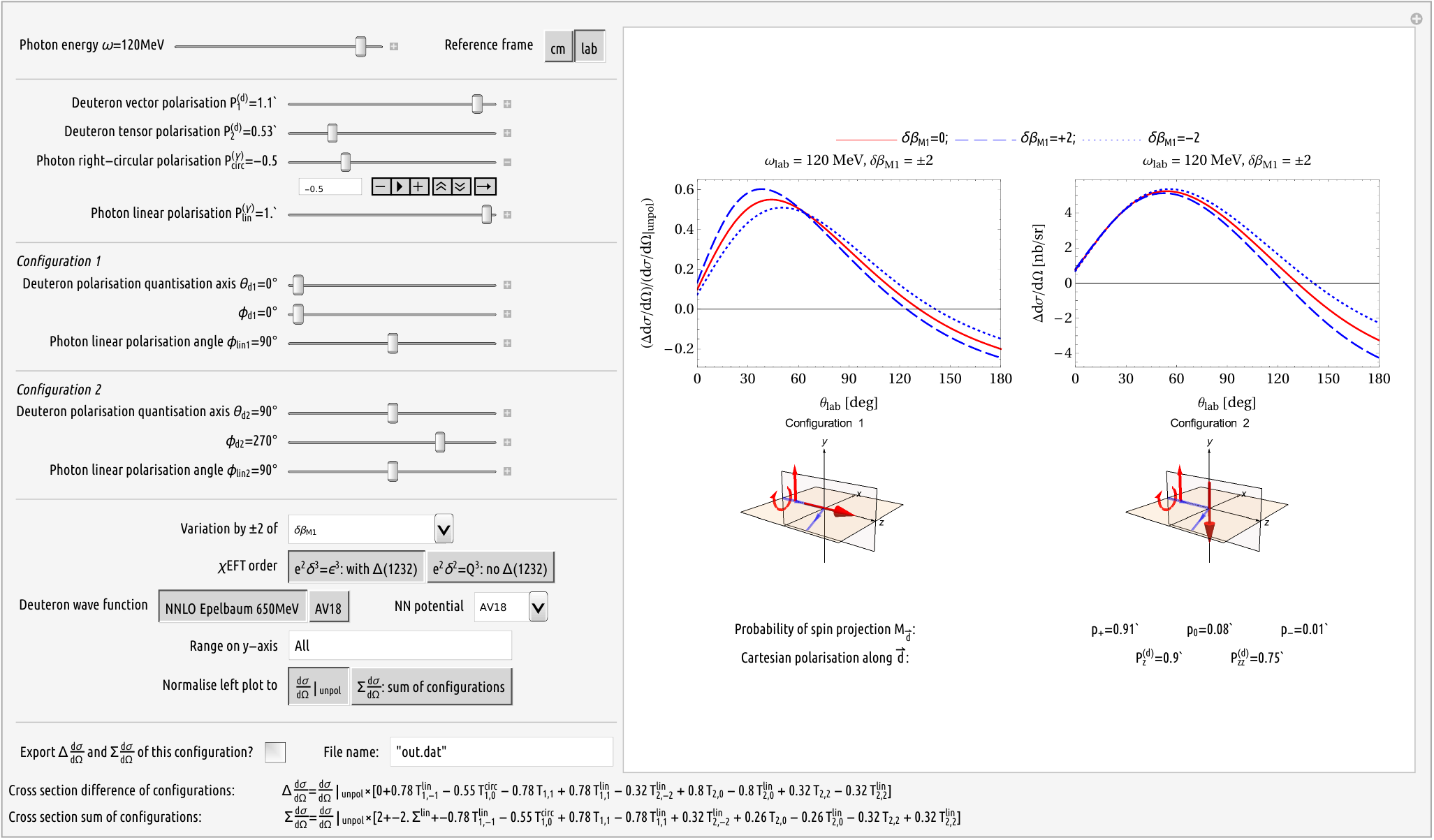}
    \caption {(Colour on-line) Screen-shot of part of the interactive
      \emph{Mathematica} notebook.}
  \label{fig:screenshot}
\end{center}
\end{figure}

To present all 17 asymmetries and their rates, plus the unpolarised cross
section, depending on $6$ dipole polarisabilities and $2$ kinematic variables
(photon energy $\omega$ and scattering angle $\theta$) in the cm and lab
frame, plus additional theoretical uncertainties and both theoretical and
experimental constraints, far exceeds what can adequately be conveyed in an
article. Here, the focus is therefore on some prominent examples. In order to
facilitate planning and analysis of experiments, the results of all
observables are available as an interactive \emph{Mathematica 9.0} notebook
from \texttt{hgrie@gwu.edu}. It contains both tables and plots of energy- and
angle-dependencies of the cross-sections, rates and asymmetries from $10$ to
about $120$~MeV, in both the cm and lab systems, including sensitivities to
varying the scalar and spin polarisabilities independently as well
as subject to the Baldin sum rule and other constraints.  Since it considers
all observables with polarised beams and/or targets, it supersedes
Ref.~\cite{Griesshammer:2010pz} which only dealt with some observables, built
in analogy to the Babusci-classification; see Sec.~\ref{sec:shuklaobs}.
Figure~\ref{fig:screenshot} shows a sample screen-shot of a cross-section
difference with user-defined beam and target polarisations.

It is finally worth re-emphasising that the purpose of this study is to
establish \emph{relative sensitivities} of Compton scattering observables on
\emph{varying} the polarisabilities~\cite{Griesshammer:2010pz}. Credible
predictions of their absolute magnitudes are only meaningful when all
systematic uncertainties are properly propagated into observables. Such errors
include: theoretical uncertainties from discarding contributions in \ChiEFT
which are higher than order $e^2\delta^3$, like including effects of the
$\Delta(1232)$ width and parameter uncertainties; uncertainties in the data
and in the Baldin Sum rule, Eq.~\eqref{eq:isoscalarBaldin}; and to a lesser
extend residual dependence on the deuteron wave-function and NN potential
used, as well as numerical uncertainties.

%%%%%%%%%%%%%%%%%%%%%%%%%%%%%%%%%%%%%%%%%%%%%%%%%%%%%%%%%%%%%%%%%%%%%%%%%%%%%%%
\subsection{Results}
\label{sec:results}

\subsubsection{Size and Sensitivity}

\begin{figure}[!hp]%[!htbp]
\begin{center}
      \includegraphics[width=\textwidth]
      {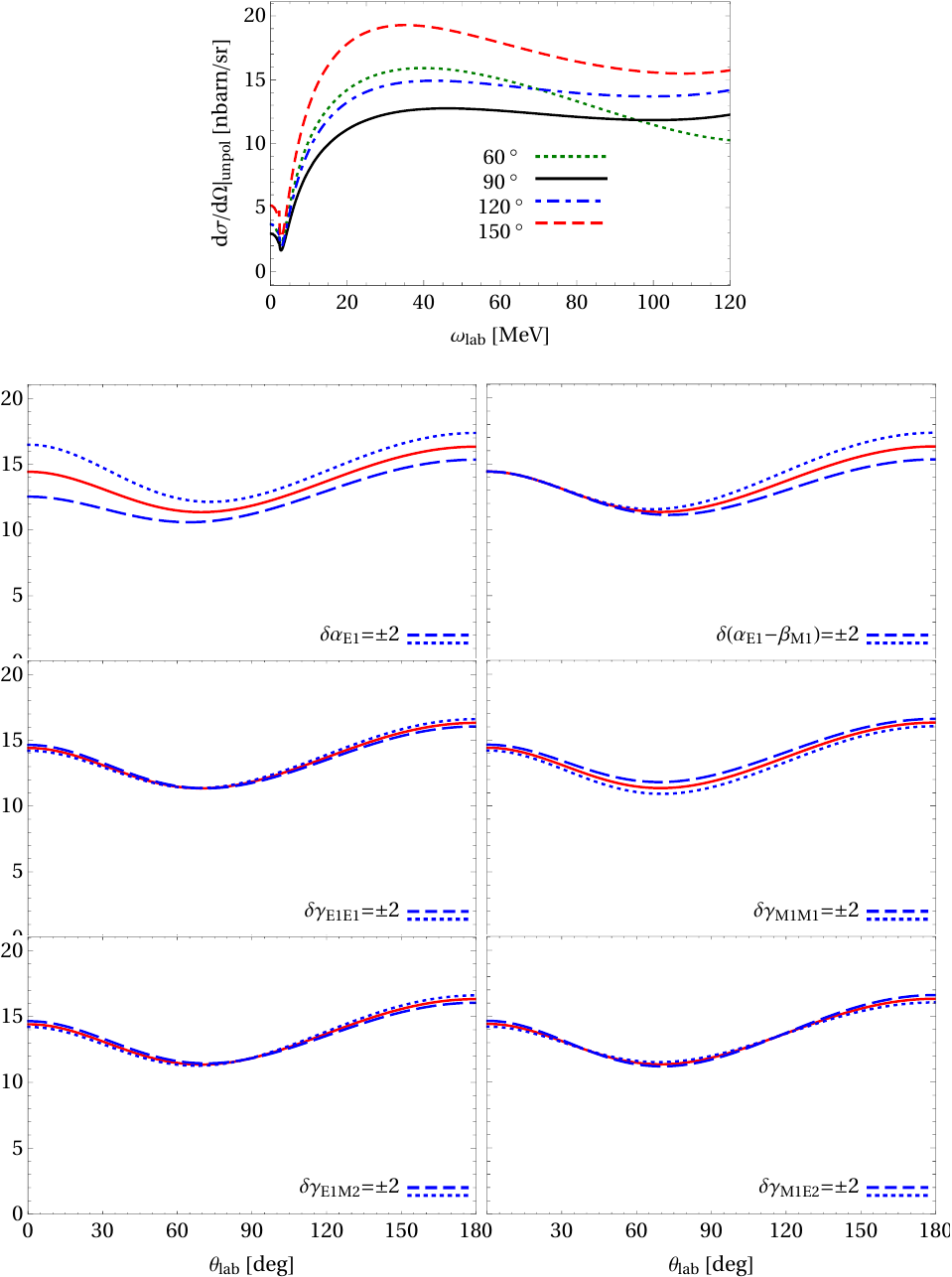}
      \caption{(Colour on-line) Unpolarised cross section
        $\dd\sigma/\dd\Omega|_\text{unpol}$ in the lab frame, in
        $\text{nbarn}/\text{sr}$. Top: energy-dependence at different angles.
        Other panels: sensitivity to varying a polarisability abouts its
        central value (\textcolor{red}{\protect\solid})
        of~Eq.~\eqref{eq:staticpols} by $+2$
        (\textcolor{blue}{\protect\longdashed}) and $-2$
        (\textcolor{blue}{\protect\dotted}) units, at
        $\omega_\text{lab}=100\;\MeV$. From top left to bottom right:
        variation of $\alphae$, $\alphae-\betam$ (constrained by the Baldin
        sum rule), $\gammaee$, $\gammamm$, $\gammaem$, $\gammame$. }
\label{fig:crosssectionvaried}
\end{center}
\end{figure}

\begin{figure}[!hp]%[!htbp]
\begin{center}
      \includegraphics[width=\textwidth]
      {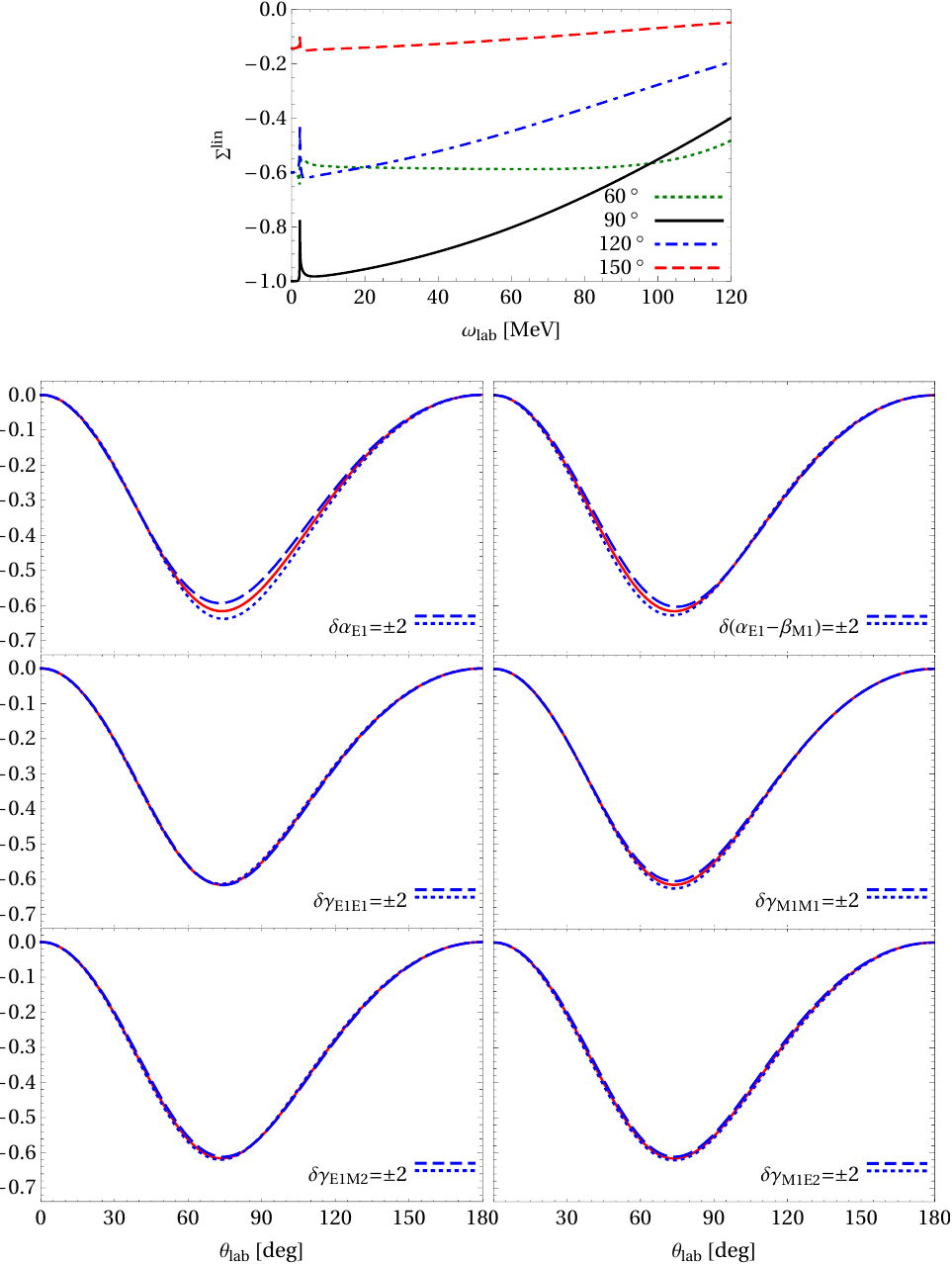}
      \caption{(Colour on-line) Beam asymmetry $\Sigma^\text{lin}$ in the
        lab frame. Top: energy-dependence at different angles.  Other panels:
        sensitivity to varying a polarisability abouts its central value
        (\textcolor{red}{\protect\solid}) of~Eq.~\eqref{eq:staticpols} by $+2$
        (\textcolor{blue}{\protect\longdashed}) and $-2$
        (\textcolor{blue}{\protect\dotted}) units, at
        $\omega_\text{lab}=100\;\MeV$. From top left to bottom right:
        variation of $\alphae$, $\alphae-\betam$, $\gammaee$, $\gammamm$,
        $\gammaem$, $\gammame$. }
\label{fig:Sigmalinvaried}
\end{center}
\end{figure}

\begin{figure}[!hp]%[!htbp]
\begin{center}
      \includegraphics[width=\textwidth]
      {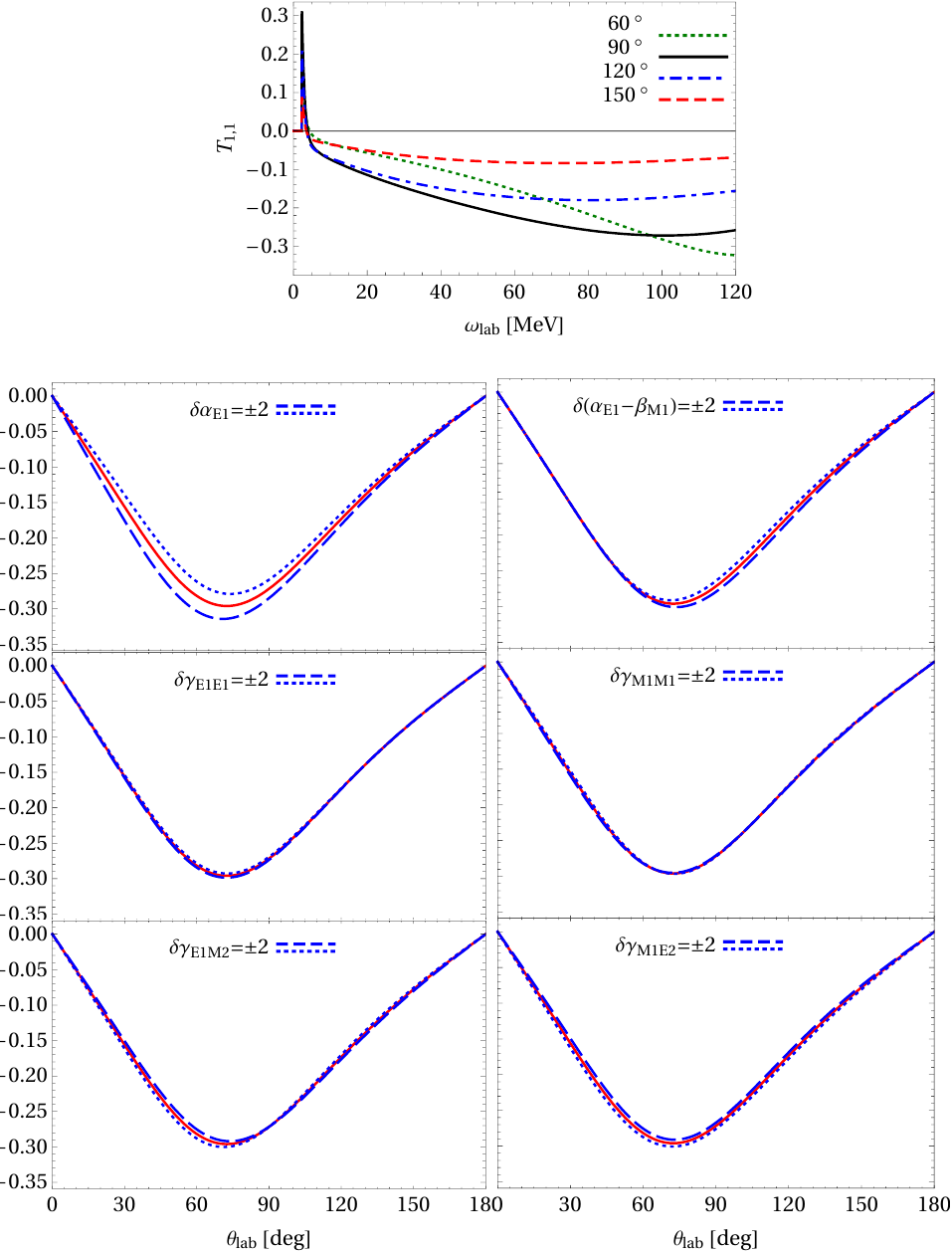}
      \caption{(Colour on-line) Vector target asymmetry $T_{11}$ (lab frame).
       See Fig.~\ref{fig:Sigmalinvaried} for notes.}
\label{fig:T11varied}
\end{center}
\end{figure}

\begin{figure}[!hp]
\begin{center}
      \includegraphics[width=\textwidth]{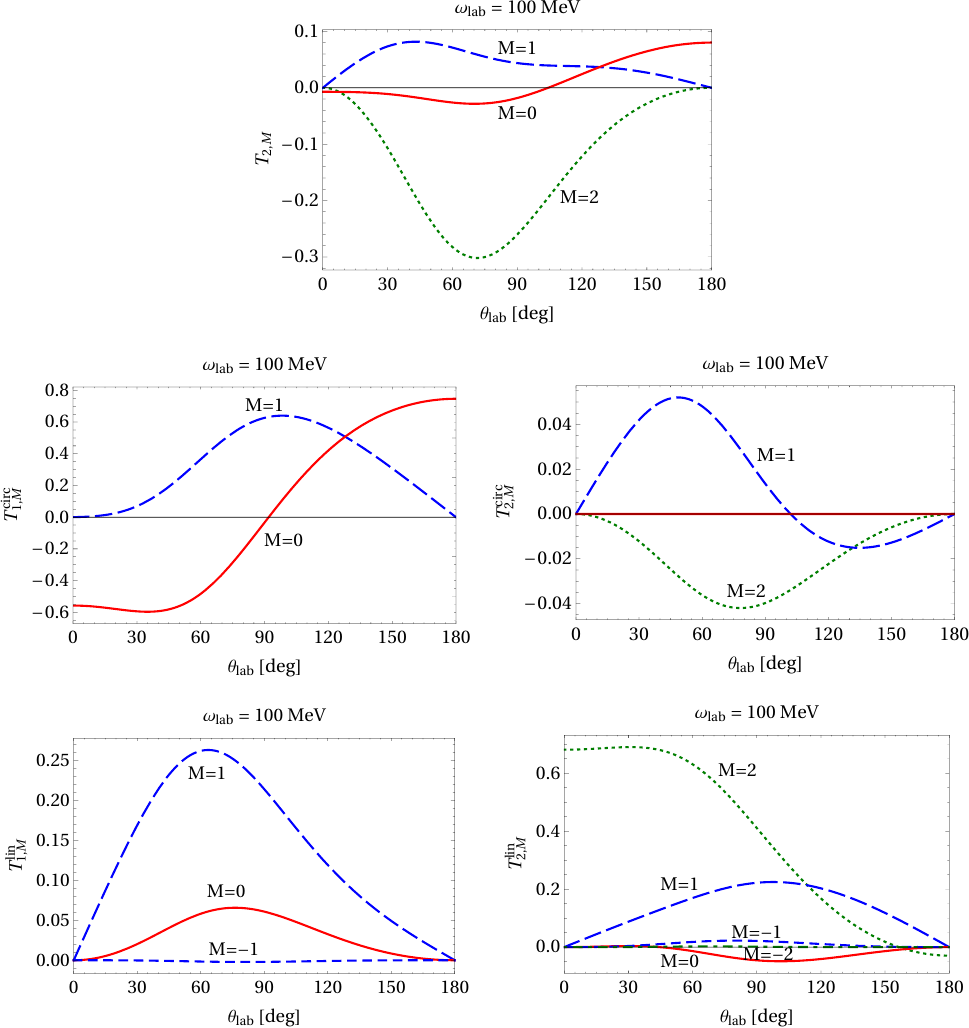}
      \caption{(Colour on-line) Comparison of the relative sizes of, from top
        left to bottom right, $T_{2M}$, $T_{1M}^\text{circ}$,
        $T_{2M}^\text{circ}$, $T_{1M}^\text{lin}$, $T_{2M}^\text{lin}$, at
        $\omega_\text{lab}=100\;\MeV$ in the lab frame, with the static
        polarisabilities given by Eq.~\eqref{eq:staticpols}.
        \textcolor{red}{\protect\solid}: $M=0$;
        \textcolor{blue}{\protect\longdashed}: $M=1$;
        \textcolor{green}{\protect\dotted}: $M=2$;
        \textcolor{blue}{\protect\shortdashed}: $M=-1$;
        \textcolor{green}{\protect\dotdashed}: $M=-2$. %
        Each panel is drawn at a different scale.}
\label{fig:compareTs}
\end{center}
\end{figure}

\begin{figure}[!hp]%[!htbp]
\begin{center}
      \includegraphics[width=\textwidth]
      {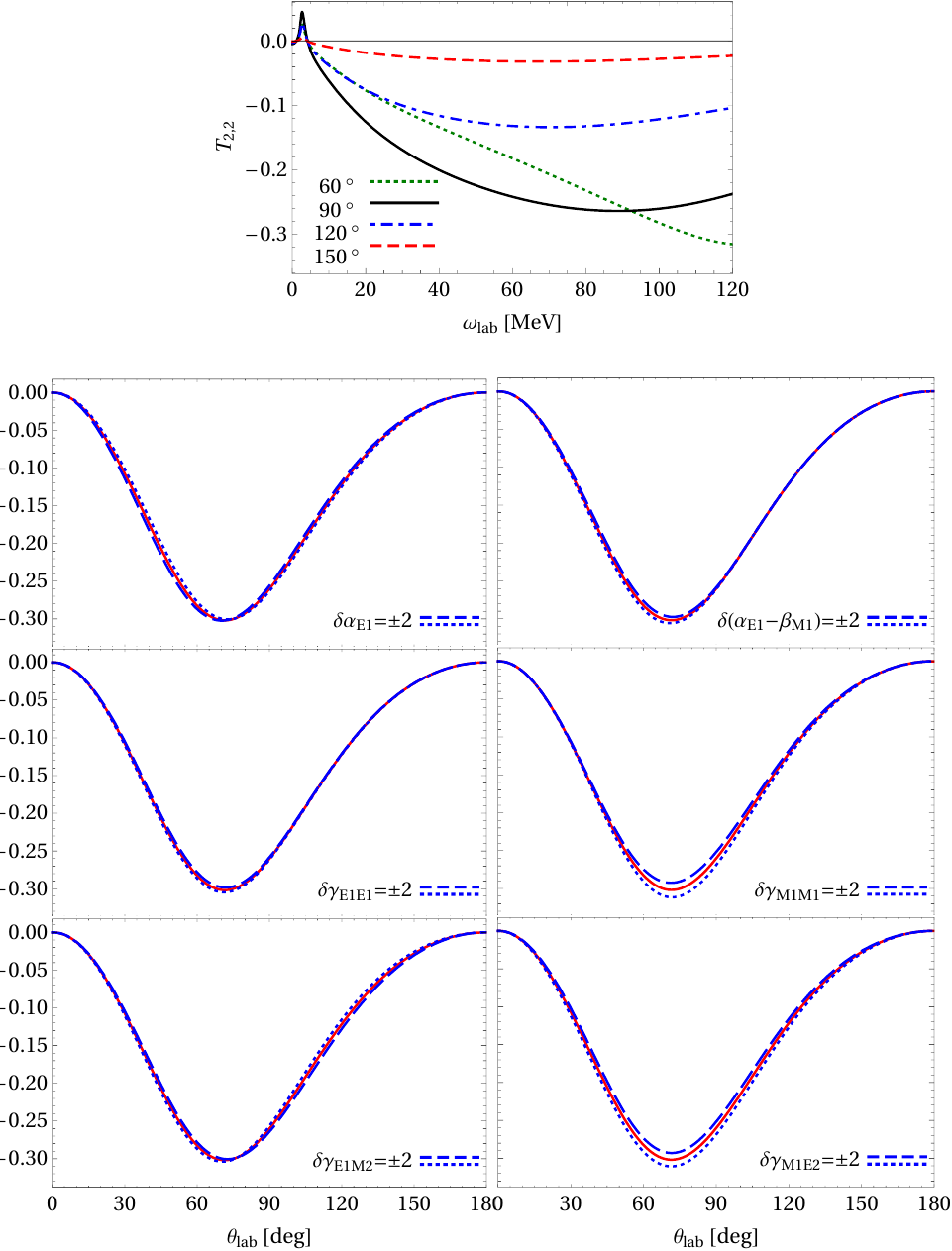}
      \caption{(Colour on-line) Tensor target asymmetry $T_{22}$ (lab frame). 
       See Fig.~\ref{fig:Sigmalinvaried} for notes.}
\label{fig:T22varied}
\end{center}
\end{figure}

\begin{figure}[!hp]%[!htbp]
\begin{center}
      \includegraphics[width=\textwidth]
      {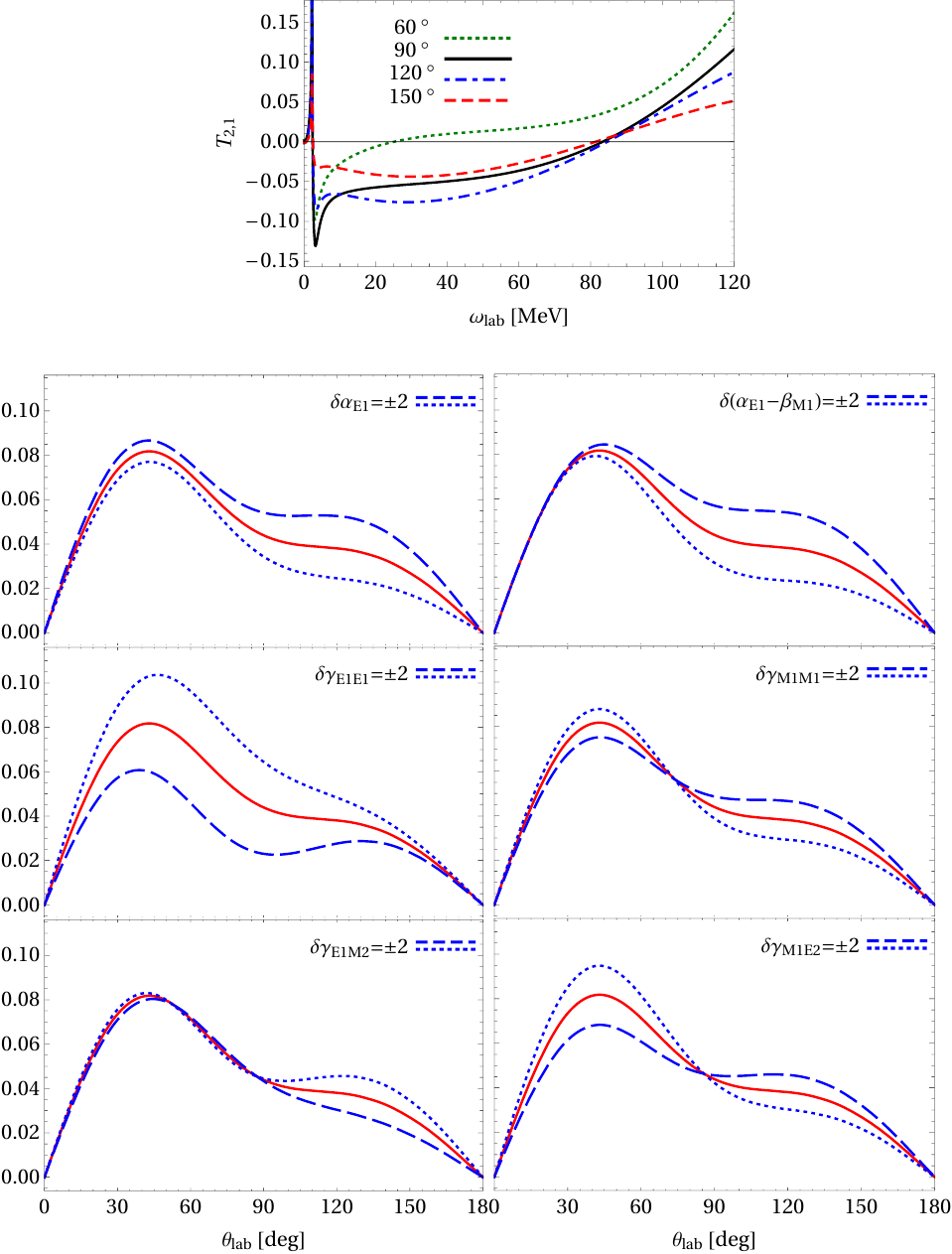}
      \caption{(Colour on-line) Tensor target asymmetry $T_{21}$ (lab frame). 
       See Fig.~\ref{fig:Sigmalinvaried} for notes.}
\label{fig:T21varied}
\end{center}
\end{figure}

\begin{figure}[!hp]%[!htbp]
\begin{center}
      \includegraphics[width=\textwidth]
      {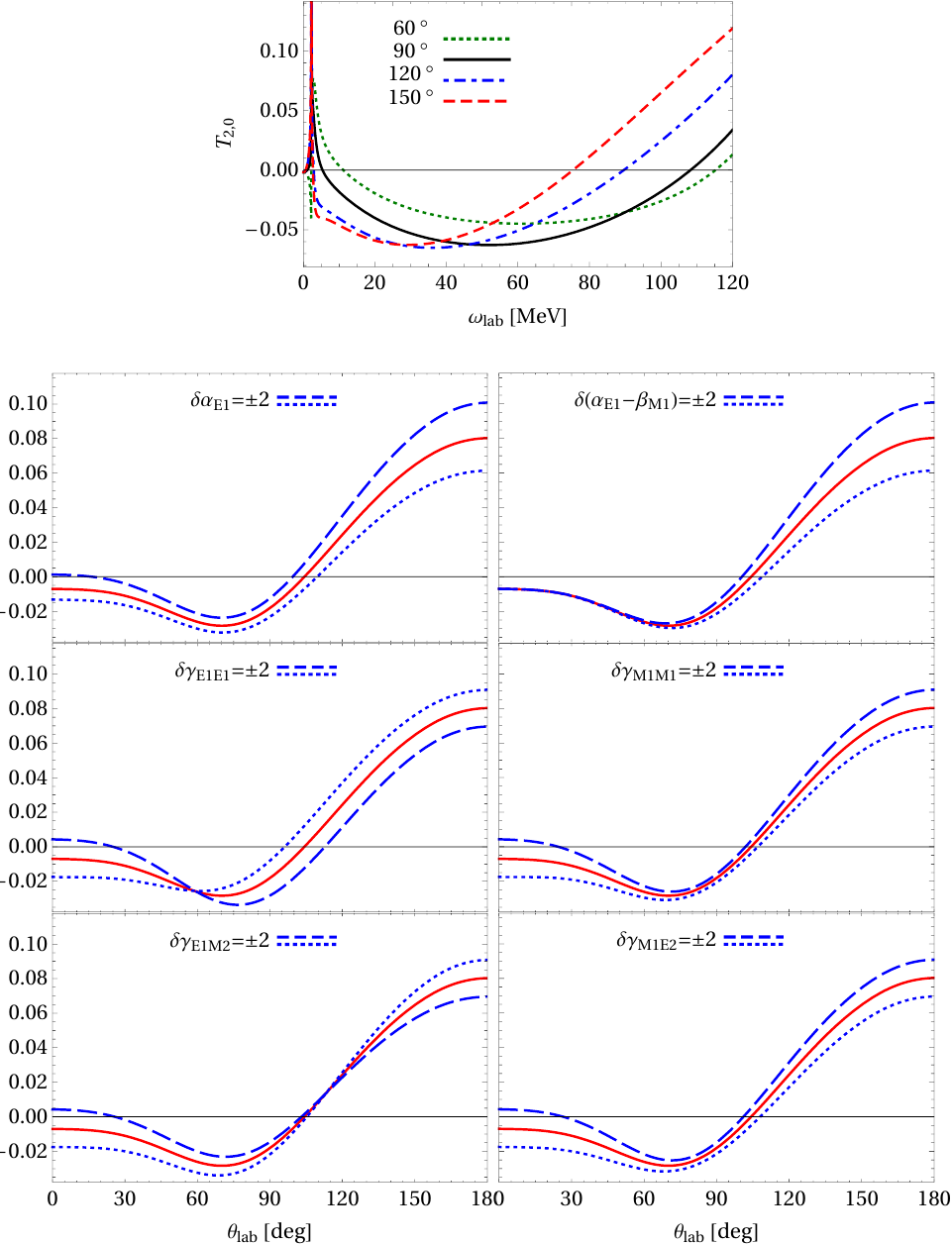}
      \caption{(Colour on-line) Tensor target asymmetry $T_{20}$ (lab frame). 
       See Fig.~\ref{fig:Sigmalinvaried} for notes.}
\label{fig:T20varied}
\end{center}
\end{figure}

\begin{figure}[!hp]%[!htbp]
\begin{center}
      \includegraphics[width=\textwidth]
      {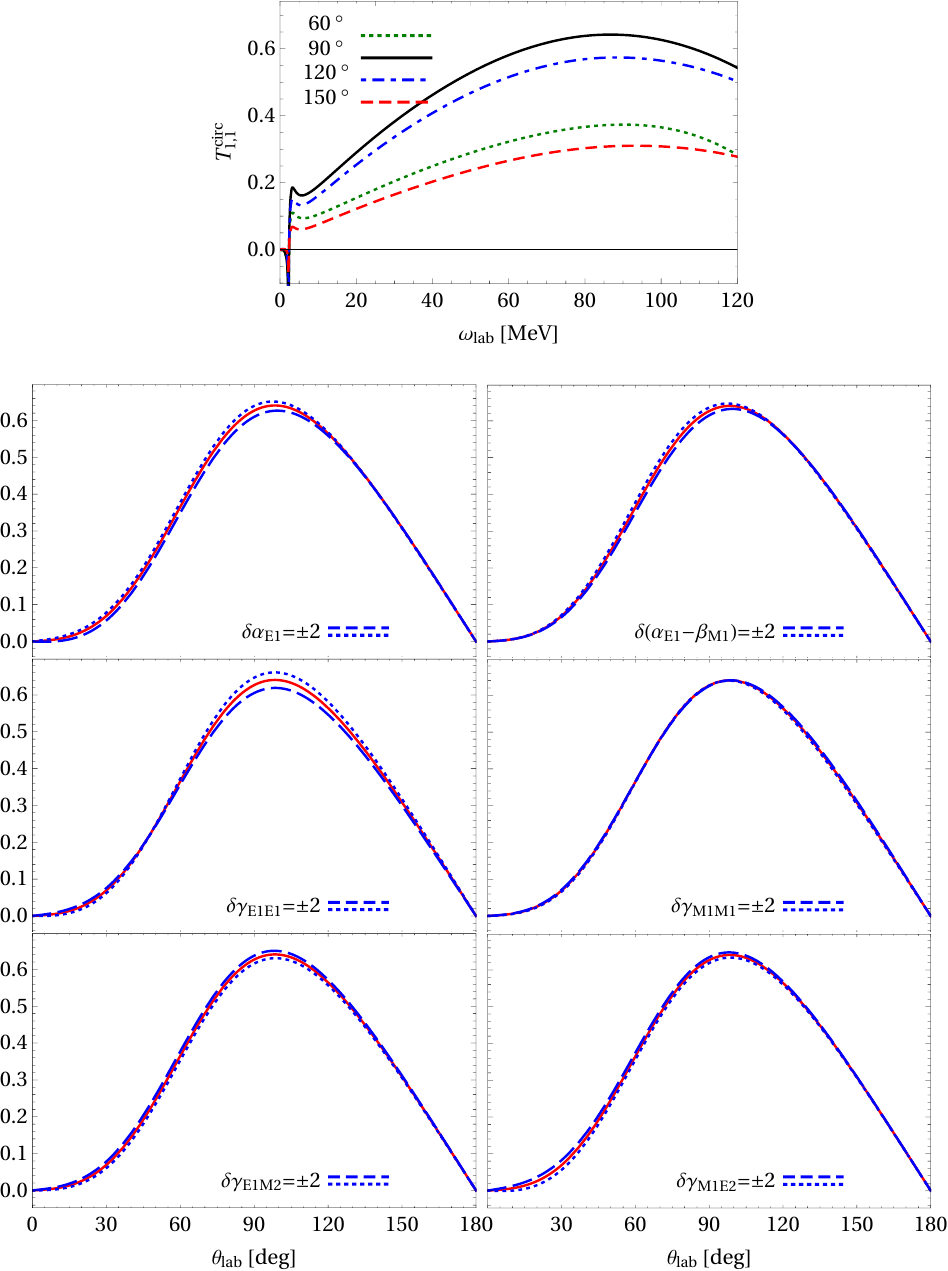}
      \caption{(Colour on-line) Double asymmetry $T_{11}^\text{circ}$ (lab frame).
       See Fig.~\ref{fig:Sigmalinvaried} for notes.}
\label{fig:Tcirc11varied}
\end{center}
\end{figure}

\begin{figure}[!hp]%[!htbp]
\begin{center}
      \includegraphics[width=\textwidth]
      {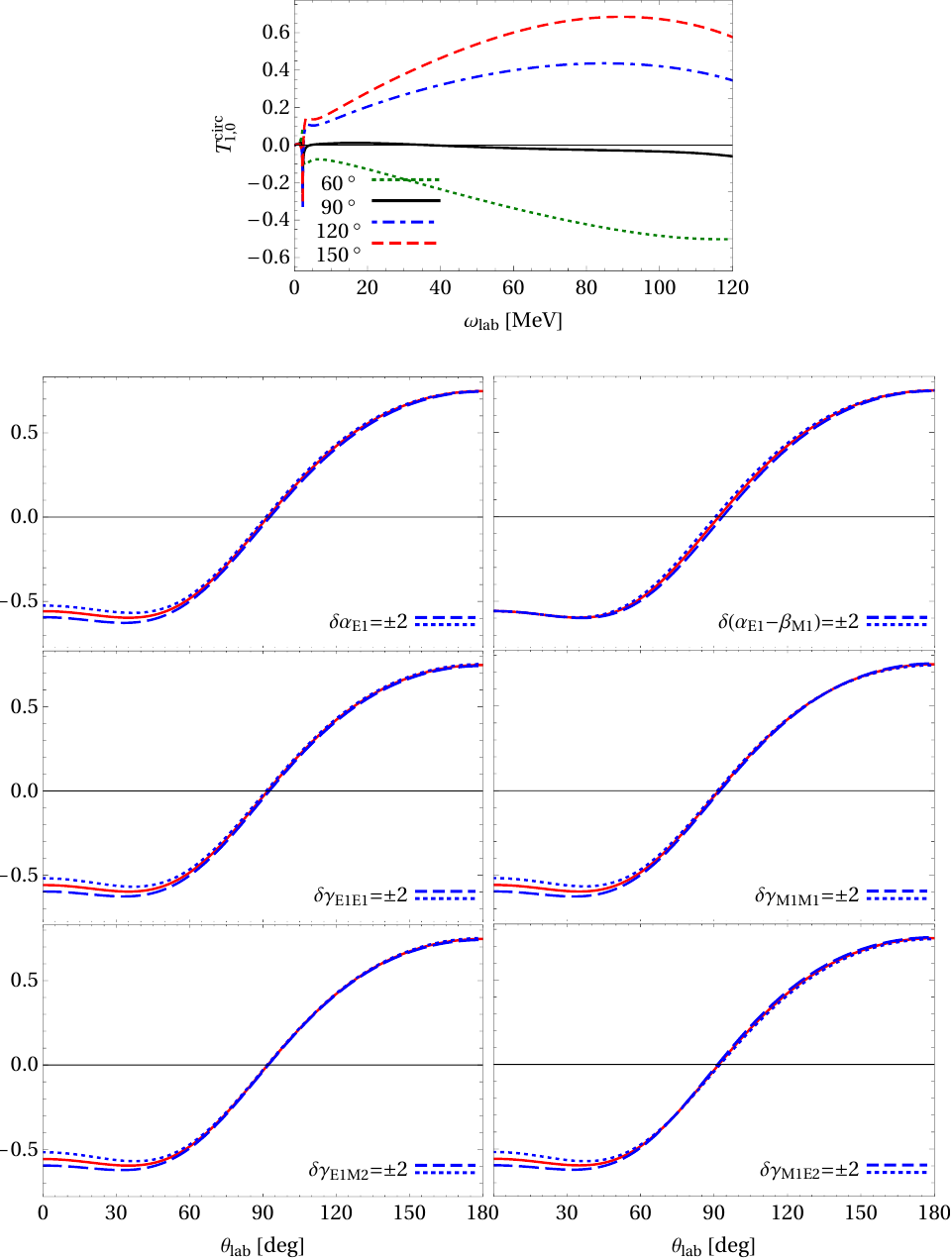}
      \caption{(Colour on-line) Double asymmetry $T_{10}^\text{circ}$ (lab frame).
       See Fig.~\ref{fig:Sigmalinvaried} for notes.}
\label{fig:Tcirc10varied}
\end{center}
\end{figure}

\begin{figure}[!hp]%[!htbp]
\begin{center}
      \includegraphics[width=\textwidth]
      {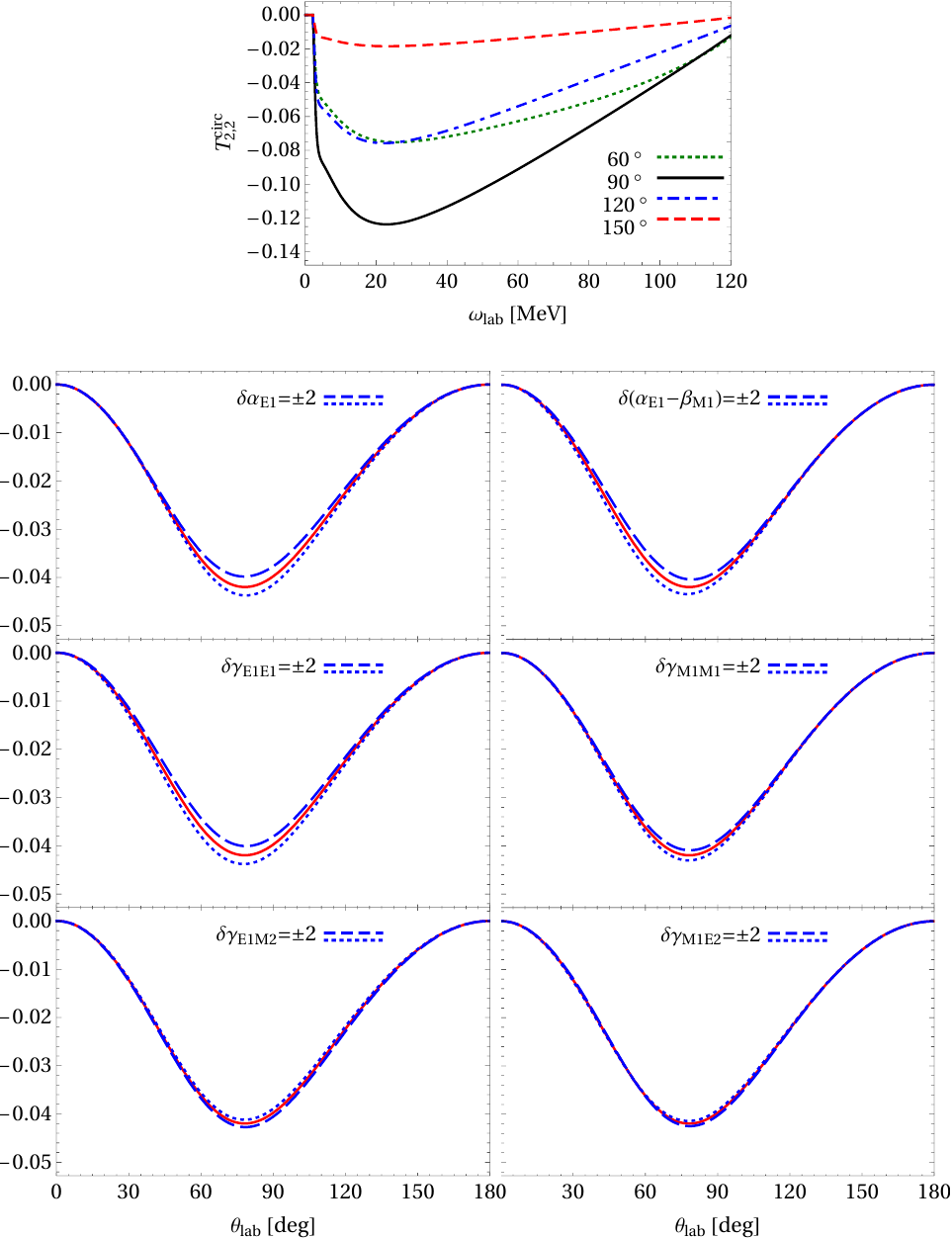}
      \caption{(Colour on-line) Double asymmetry $T_{22}^\text{circ}$ (lab frame).
       See Fig.~\ref{fig:Sigmalinvaried} for notes.}
\label{fig:Tcirc22varied}
\end{center}
\end{figure}

\begin{figure}[!hp]%[!htbp]
\begin{center}
      \includegraphics[width=\textwidth]
      {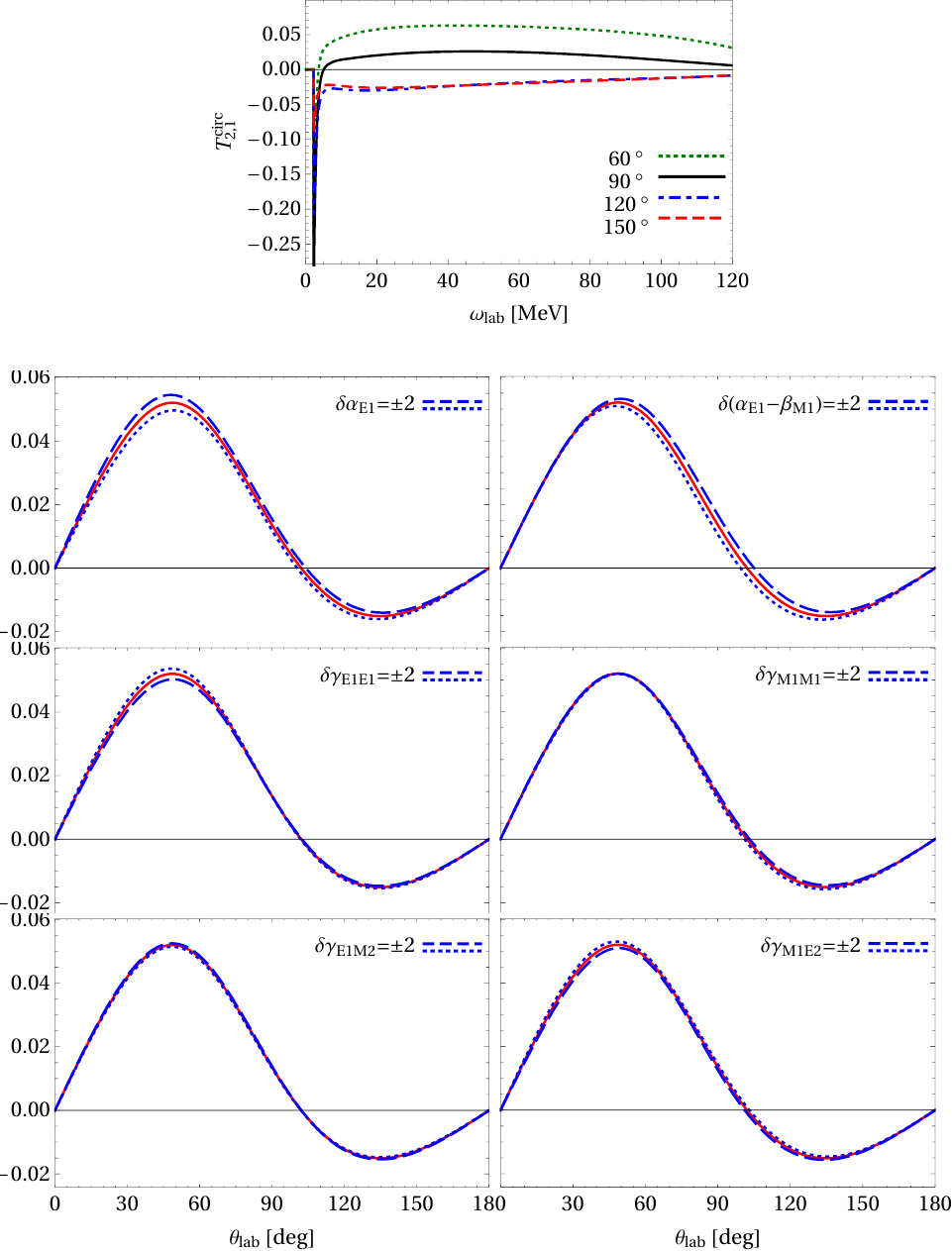}
      \caption{(Colour on-line) Double asymmetry $T_{21}^\text{circ}$ (lab frame).
       See Fig.~\ref{fig:Sigmalinvaried} for notes.}
\label{fig:Tcirc21varied}
\end{center}
\end{figure}

\begin{figure}[!hp]%[!htbp]
\begin{center}
      \includegraphics[width=\textwidth]
      {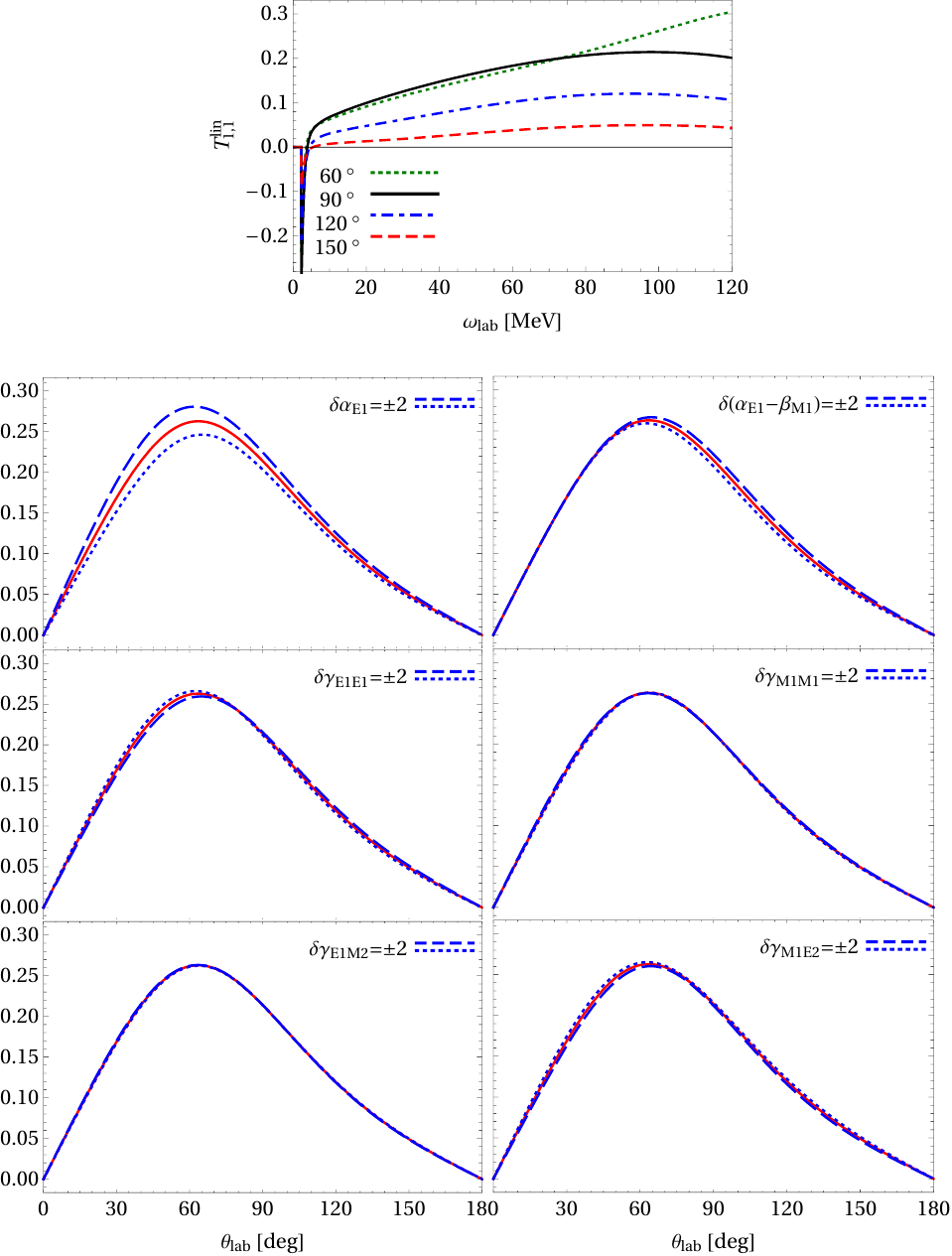}
      \caption{(Colour on-line) Double asymmetry $T_{11}^\text{lin}$ (lab frame).
       See Fig.~\ref{fig:Sigmalinvaried} for notes.}
\label{fig:Tlin11varied}
\end{center}
\end{figure}

\begin{figure}[!hp]%[!htbp]
\begin{center}
      \includegraphics[width=\textwidth]
      {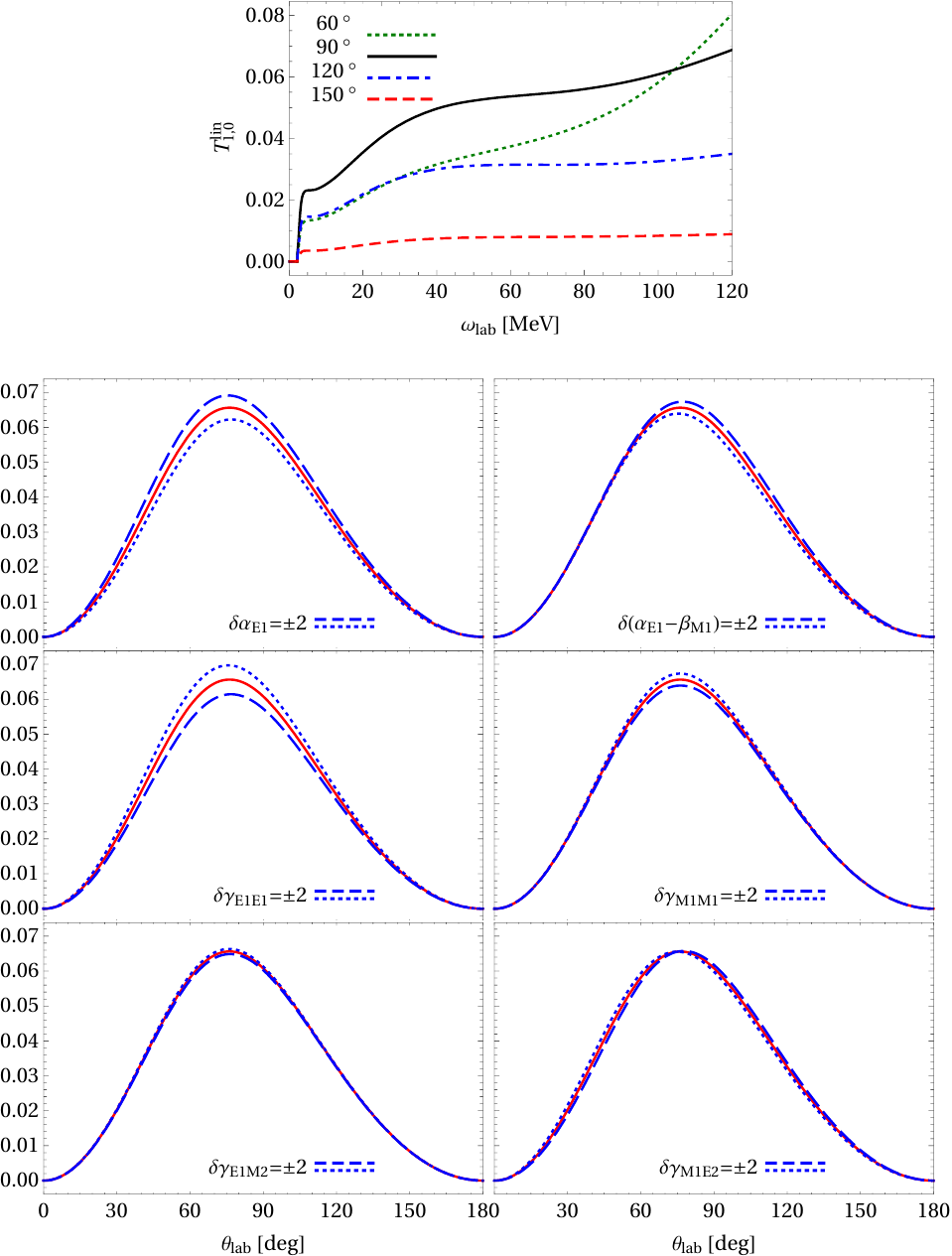}
      \caption{(Colour on-line) Double asymmetry $T_{10}^\text{lin}$ (lab frame).
       See Fig.~\ref{fig:Sigmalinvaried} for notes.}
\label{fig:Tlin10varied}
\end{center}
\end{figure}

\begin{figure}[!hp]%[!htbp]
\begin{center}
      \includegraphics[width=\textwidth]
      {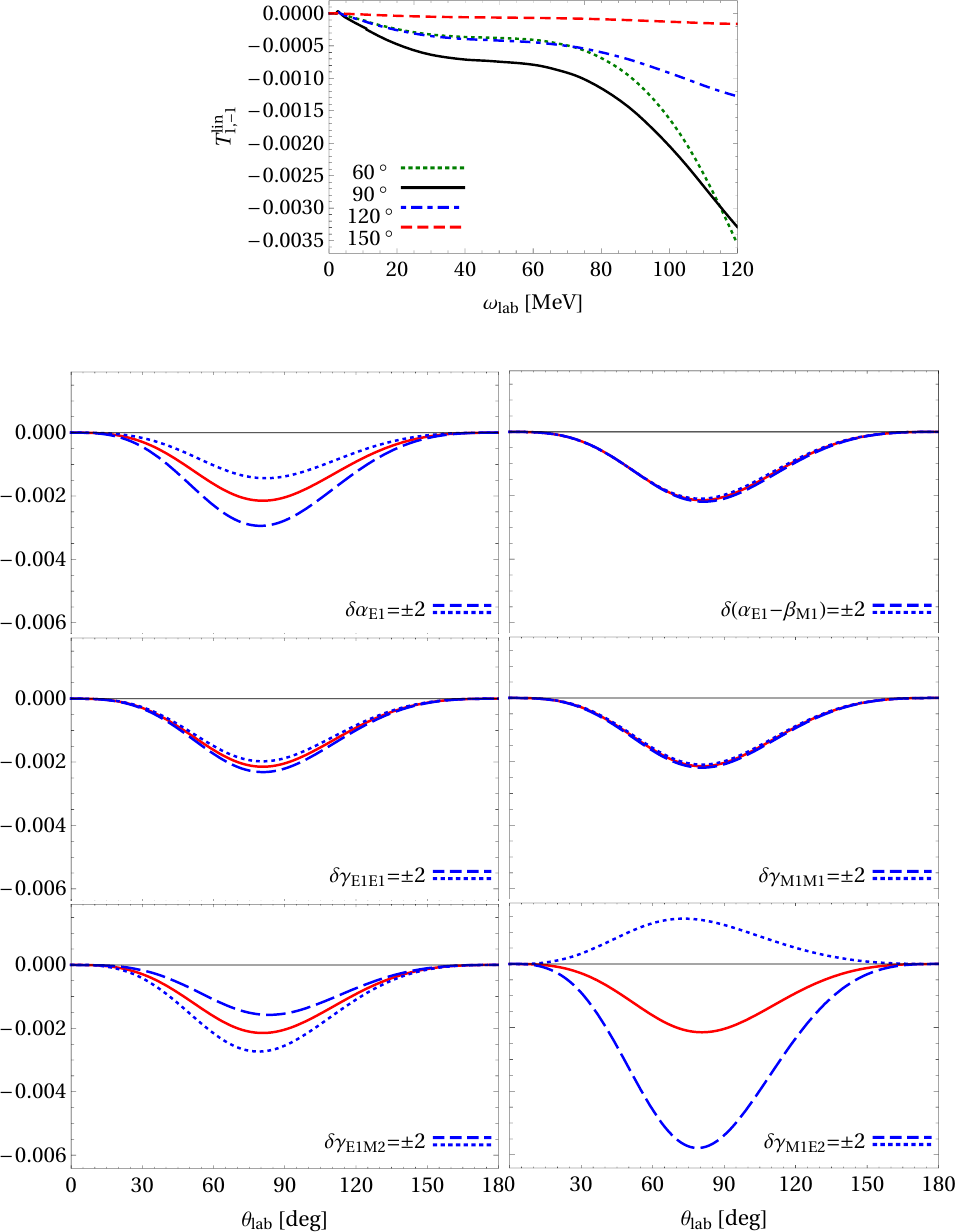}
      \caption{(Colour on-line) Double asymmetry $T_{1,-1}^\text{lin}$ (lab frame).
       See Fig.~\ref{fig:Sigmalinvaried} for notes.}
\label{fig:Tlin1-1varied}
\end{center}
\end{figure}

\begin{figure}[!hp]%[!htbp]
\begin{center}
      \includegraphics[width=\textwidth]
      {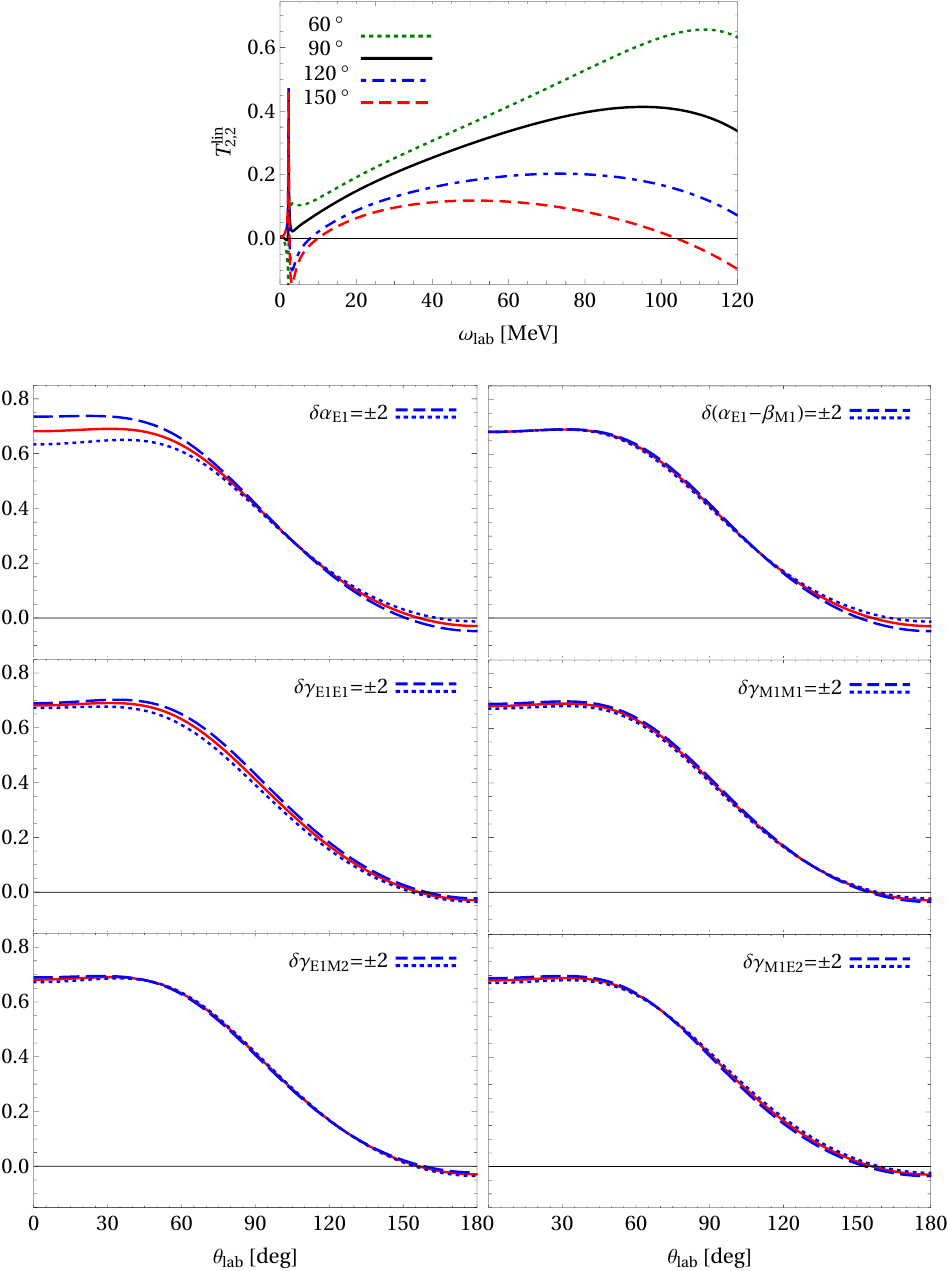}
      \caption{(Colour on-line) Double asymmetry $T_{22}^\text{lin}$ (lab frame).
       See Fig.~\ref{fig:Sigmalinvaried} for notes.}
\label{fig:Tlin22varied}
\end{center}
\end{figure}

\begin{figure}[!hp]%[!htbp]
\begin{center}
      \includegraphics[width=\textwidth]
      {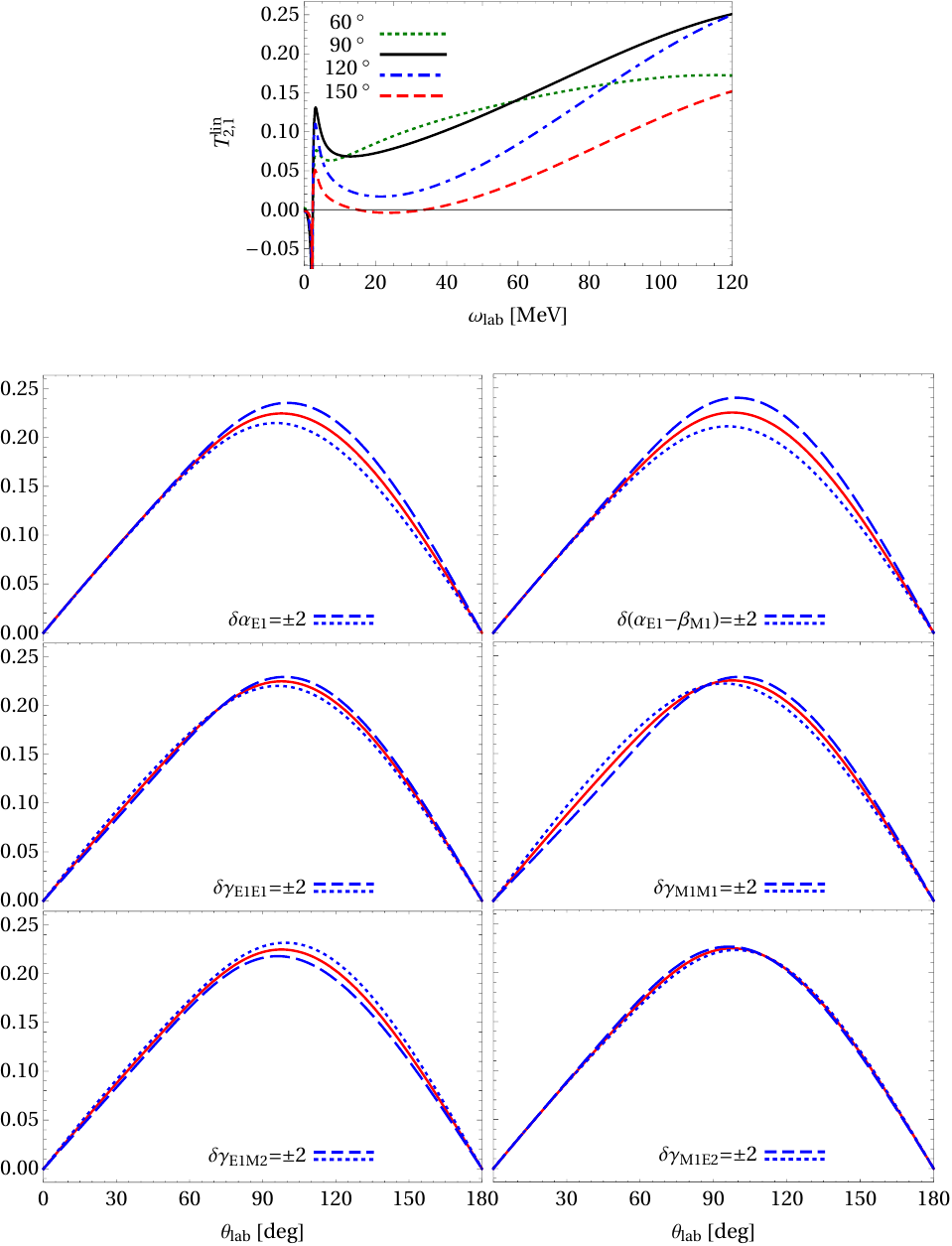}
      \caption{(Colour on-line) Double asymmetry $T_{21}^\text{lin}$ (lab frame).
       See Fig.~\ref{fig:Sigmalinvaried} for notes.}
\label{fig:Tlin21varied}
\end{center}
\end{figure}

\begin{figure}[!hp]%[!htbp]
\begin{center}
      \includegraphics[width=\textwidth]
      {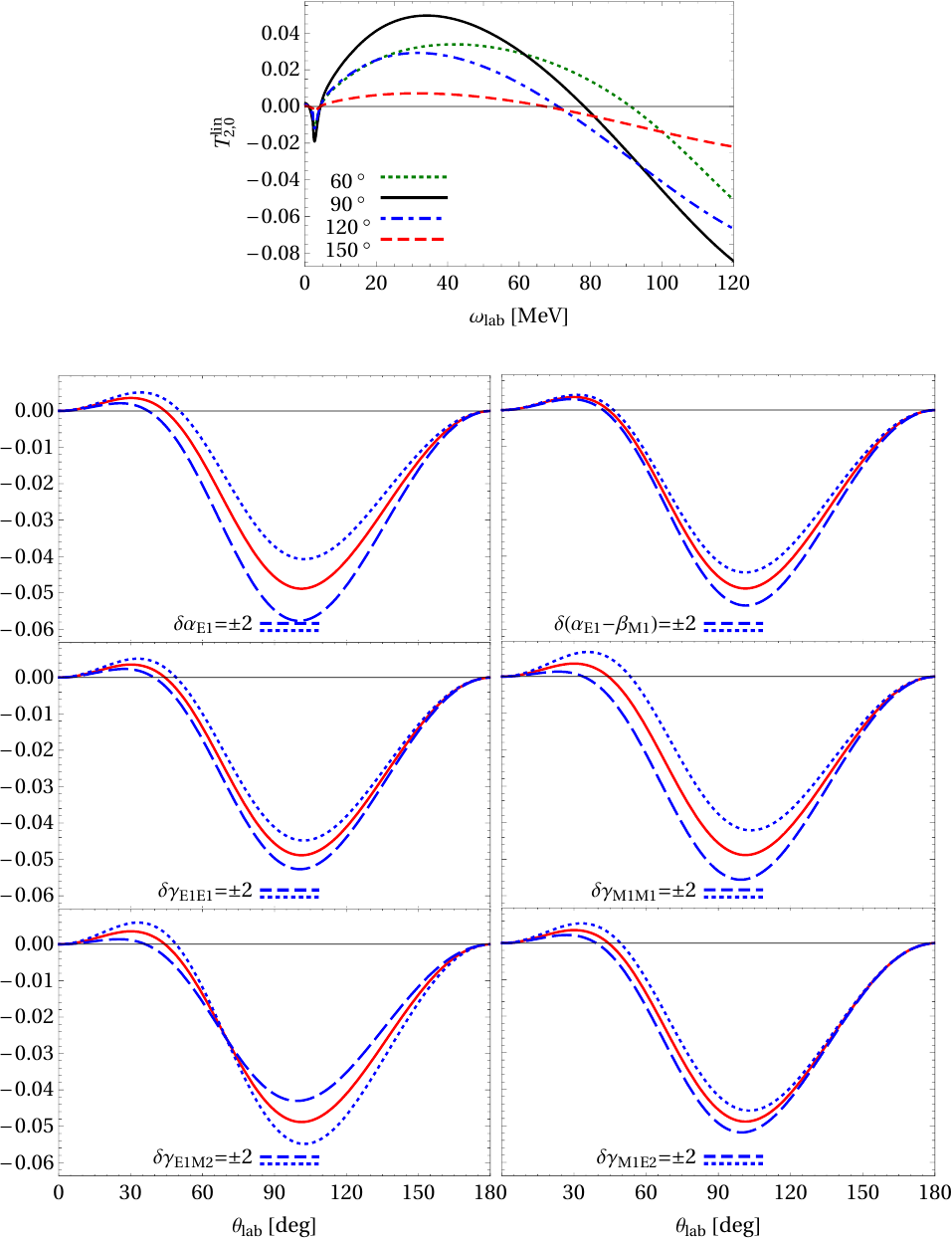}
      \caption{(Colour on-line) Double asymmetry $T_{20}^\text{lin}$ (lab frame).
       See Fig.~\ref{fig:Sigmalinvaried} for notes.}
\label{fig:Tlin20varied}
\end{center}
\end{figure}

\begin{figure}[!hp]%[!htbp]
\begin{center}
      \includegraphics[width=\textwidth]
      {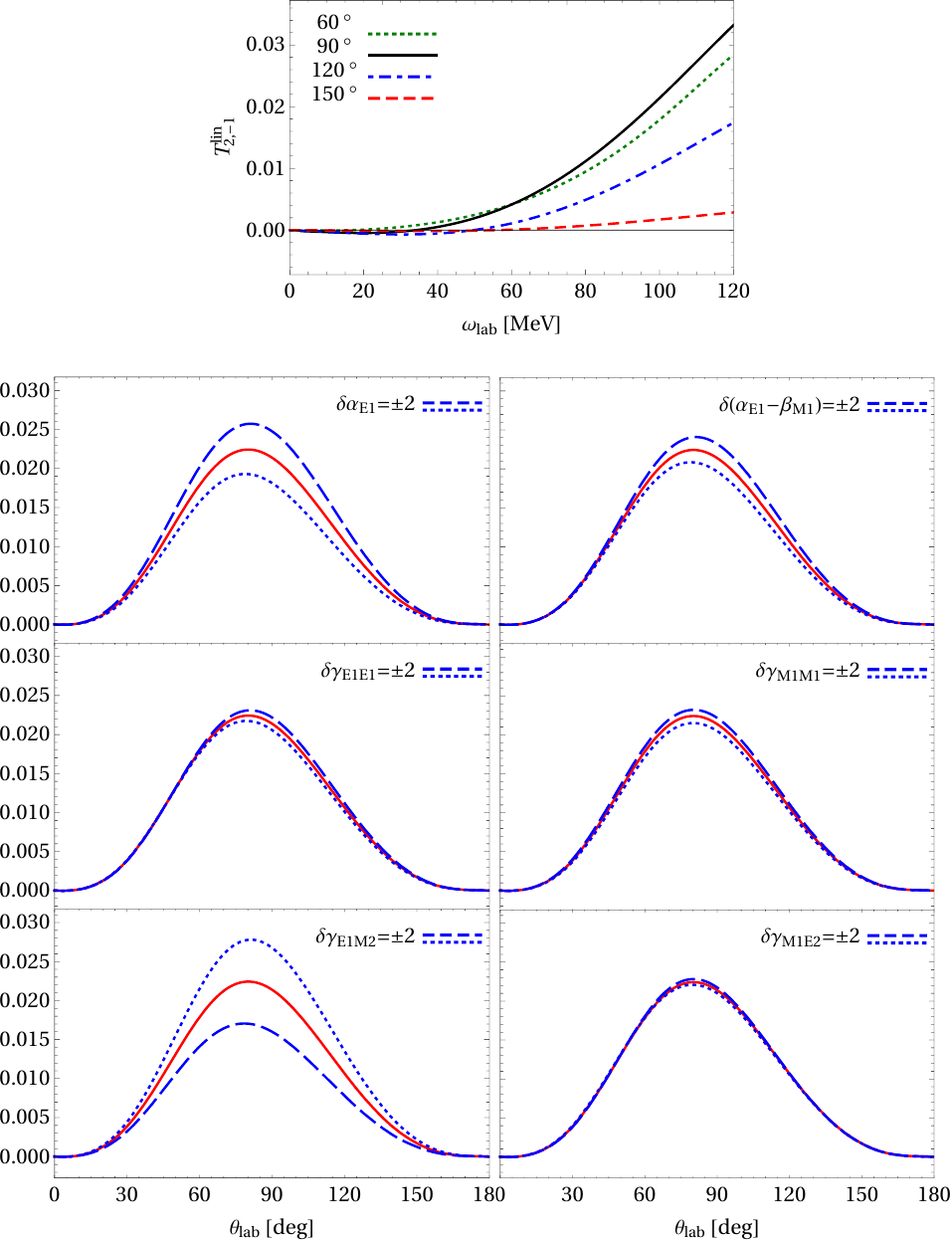}
      \caption{(Colour on-line) Double asymmetry $T_{2,-1}^\text{lin}$ (lab frame).
       See Fig.~\ref{fig:Sigmalinvaried} for notes.}
\label{fig:Tlin2-1varied}
\end{center}
\end{figure}

\begin{figure}[!hp]%[!htbp]
\begin{center}
      \includegraphics[width=\textwidth]
      {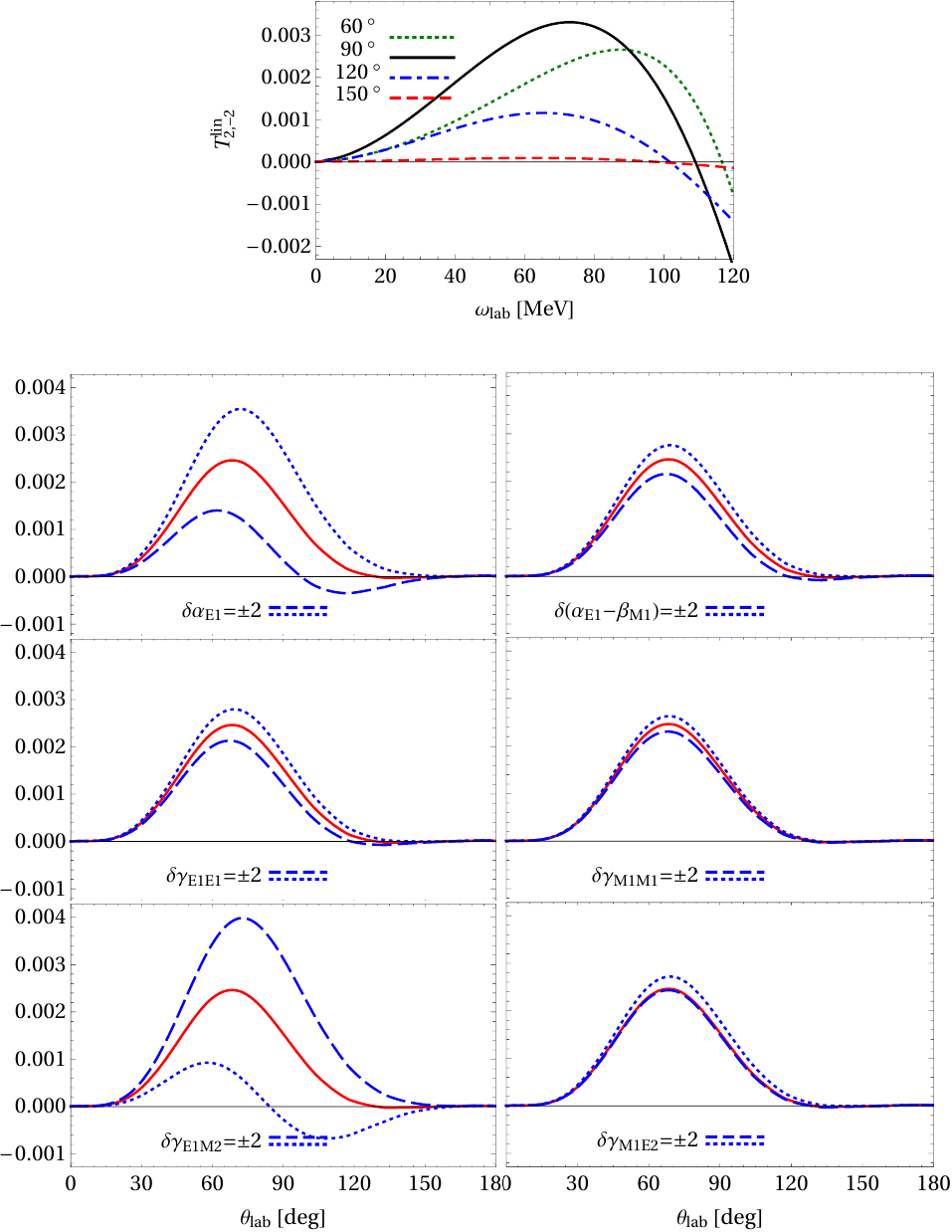}
      \caption{(Colour on-line) Double asymmetry $T_{2,-2}^\text{lin}$ (lab frame).
       See Fig.~\ref{fig:Sigmalinvaried} for notes.}
\label{fig:Tlin2-2varied}
\end{center}
\end{figure}

Figures~\ref{fig:crosssectionvaried} to \ref{fig:Tlin2-2varied} present the
\ChiEFT results of an $\calO(e^2\delta^3)$-calculation, with dynamical
$\Delta(1232)$ and NN-rescattering. Let us concentrate on the sensitivity
to the polarisabilities at one representative energy in the (experimentally
most relevant) lab-frame. With an eye on parameters at \HIGS, MAXlab, MAMI and
possible future high-luminosity accelerators like MESA~\cite{nocit}, a beam
energy of $\omega_\text{lab}=100\;\MeV$ seems appropriate. Staying below the
pion-production threshold avoids experimental and theoretical complications.

Since the asymmetries differ by $3$ orders of magnitude, one should keep in
mind changes of scale between plots of different observables. Comparing them
is simplified by plots of $T_{2M}$, $T^\text{circ}_{1M}$,
$T^\text{circ}_{2M}$, $T^\text{lin}_{1M}$ and $T^\text{lin}_{2M}$, each for
the different non-trivial values of $M$ at $\omega_\text{lab}=100\;\MeV$.
With magnitudes of up to $0.7$, the largest asymmetries are
$\Sigma^\text{lin}$, $T^\text{circ}_{1M}$ and $T^\text{lin}_{22}$, followed by
magnitudes of about $[0.06\dots0.3]$ for $T_{JM}$, $T^\text{lin}_{1(1,0)}$ and
$T^\text{lin}_{2(1,0)}$.  The order of magnitude of $T^\text{circ}_{2M}$,
$T^\text{lin}_{10}$ and $T^\text{lin}_{2,-1}$ is $10^{-2}$, and that of
$T^\text{lin}_{1,-1}$ and $T^\text{lin}_{2,-2}$ even $10^{-3}$, providing
considerable experimental challenges. The observables $T^\text{lin}_{JM}$ show
a clear hierarchy, with sizes increasing substantially towards the most
positive $M$-values at given $J$.

The top panel of each single-observable plot,
Figs.~\ref{fig:crosssectionvaried} to \ref{fig:T11varied} and
\ref{fig:T22varied} to \ref{fig:Tlin2-2varied}, shows the energy-dependence of
each observable at four scattering angles
$\theta_\text{lab}\in\{60^\circ;90^\circ; 120^\circ; 150^\circ\}$. In each
case, the deuteron breakup point at $\omega_\text{lab}\approx3\;\MeV$ is
clearly visible. Only $T^\text{circ}_{22}$ and $T^\text{lin}_{2,-2}$
significantly decrease with increasing photon energy, but $T_{2(1,0)}$ and
$T^\text{lin}_{20}$ change sign around $90\;\MeV$. All observables which are
zero below the first threshold, Eq.~\eqref{eq:zerobelow}, grow rapidly in
magnitude above it -- in the case of $T_{11}$ and $T^\text{lin}_{11}$ even to
$\approx\pm0.2$ at $100\;\MeV$.

Sensitivity on the nucleon polarisabilities grows as expected with increasing
photon energy. In the lower panels of Figs.~\ref{fig:crosssectionvaried} to
\ref{fig:T11varied} and \ref{fig:T22varied} to \ref{fig:Tlin2-2varied}, two
plots show the sensitivity to $\alphae$ and the combination $\alphae-\betam$
when the Baldin sum rule constraint is used. This of course also allows one to
assess where variations of $\betam$ are (anti-)correlated to those of
$\alphae$. The other 4 panels describe variations of the spin
polarisabilities, without imposing additional constraints. Within one
observable, all sensitivities are of course plotted on the same scale.

Plots of the unpolarised cross section, Fig.~\ref{fig:crosssectionvaried}, are
included for quick rate-estimates. Its overall size is dramatically affected
by a variation of $\alphae$, its backward angles by that of $\alphae-\betam$,
and there is only minor sensitivity on the spin polarisabilities.

The beam asymmetry $\Sigma^\text{lin}$ shows a mildly different angular
dependence on $\alphae$ and $\betam$, possibly allowing for extractions. That
sensitivity to the other polarisabilities is small, had already been
demonstrated in a \ChiEFT variant without dynamical $\Delta(1232)$ in
Refs.~\cite{Choudhury:2004yz,ShuklaThesis}. Delta-effects affect this variable
only minimally.

In a future world of high-accuracy experiments with well-controlled systematic
experimental uncertainties, high luminosities and $100\%$ beam and target
polarisations, an ideal observable should be very sensitive to one
polarisability, while being near-insensitive to all others. For $\alphae$,
this singles out $T_{11}$ (Fig.~\ref{fig:T11varied}), $T^\text{lin}_{11}$
(Fig.~\ref{fig:Tlin11varied}) and $T^\text{lin}_{22}$
(Fig.~\ref{fig:Tlin22varied}); for $\gammaee$, $T^\text{circ}_{11}$
(Fig.~\ref{fig:Tcirc11varied}).  When one takes $\alphae$ and $\betam$ to be
know sufficiently well that the influence on varying them can be neglected,
then $T^\text{circ}_{11}$ (Fig.~\ref{fig:Tcirc11varied}),
$T^\text{circ}_{2(2,1)}$ (Figs.~\ref{fig:Tcirc22varied} and
\ref{fig:Tcirc21varied}) and $T^\text{lin}_{10}$ (Fig.~\ref{fig:Tlin10varied})
are dominated by sensitivity to $\gammaee$ only. 
Curiously, $T^\text{lin}_{1,-1}$ (Fig.~\ref{fig:Tlin1-1varied}) is
near-exclusively sensitive to the mixed spin polarisability $\gammame$, and
both $T^\text{lin}_{2,-2}$ (Fig.~\ref{fig:Tlin2-2varied}) and
$T^\text{lin}_{2,-1}$ (Fig.~\ref{fig:Tlin2-1varied}) to its partner $\gammaem$
-- albeit all three are very small.

Alternatively, different angular dependencies can be used to dis-entangle two
polarisabilities from the same observable; see e.g.~$T_{21}$ for $\gammaee$
and $\gammame$ (Fig.~\ref{fig:T21varied}) and -- to a lesser extend --
$T^\text{lin}_{20}$ for $\gammamm$ and $\gammaem$
(Fig.~\ref{fig:Tlin20varied}). Keeping in mind that none of the tensor
observables have an analogue in Compton scattering off the nucleon, such an
augmentation is absent in the one-nucleon case. It appears that mixed
polarisabilities are much better accessible in scattering from the deuteron.
The photon quadrupole coupling to one nucleon ($M2$ in $\gammaem$ and $E2$ in
$\gammame$) seems to be enhanced by the $D$ wave components of the deuteron
wave function and pion-exchange current, Fig.~\ref{fig:twonucleon} (a) and
(c). One may thus speculate that determinations of $\gammaem$ and $\gammame$
will first appear from deuteron data -- if the necessary accuracy can be
reached for these small asymmetries.

References~\cite{Maximon:1989zz,Griesshammer:2010pz,Griesshammer:2012we} have
argued in detail that sensitivity to a specific polarisability can be
maximised or switched off by considering particular target-beam combinations
at particular angles. To that end, one either maximises the scalar products
between photon polarisations $\vec{\epsilon}$, $\vec{\epsilon}^\prime$, photon
momenta $\kv$, $\kv^\prime$ and nucleon spin $\vec{\sigma}$, or one chooses
some vectors to be orthogonal or parallel, rendering the associated (scalar or
vector) products zero. Many of these ``zero sensitivity points'' are preserved
when the relative motion of the $\gamma N$ cm system inside the deuteron is
taken into account. In some cases, the deuteron effect lifts the zero, but
only barely, since the nucleons are predominantly in a relative $S$ wave,
while $D$ wave contributions (also from pion-exchange currents,
Fig.~\ref{fig:twonucleon} (a/c)) are suppressed. Relativistic boost effects
are small at the energies considered~\cite{Beane:2004ra}. Examples include the
following insensitivities (angles in cm frame): $T_{2(2,0)}$ to $\betam$ at
$90^\circ$; $T_{20}$ to $\gammaee$ at $60^\circ$ and to $\gammaem$ at
$120^\circ$; $T_{21}$ to $\gammaem$ and $\gammame$ at $90^\circ$;
$T^\text{circ}_{21}$ to $\gammaee$ at $90^\circ$; and $T^\text{lin}_{21}$ to
$\gammamm$ at $90^\circ$.

A good example of undesired correlations between variations of different
polarisabilities is $T^\text{circ}_{22}$ (Fig.~\ref{fig:Tcirc22varied}), where
angular dependencies and magnitudes of changing $\alphae$ and $\gammaee$ are
near-identical. $T^\text{circ}_{10}$ (Fig.~\ref{fig:Tcirc10varied}) is
near-equally sensitive to all dipole polarisabilities.

Applying these criteria and assuming that $\alphae$ and $\betam$ are known,
the following observables could therefore provide an experimentally realistic
but challenging complete set from which to cleanly determine the isoscalar
spin polarisabilities: $T^\text{circ}_{11}$ for $\gammaee$ (variation by
$\pm2$ translates into $\pm5\%$ of an asymmetry magnitude of about $0.7$),
followed by angular dependence of $T^\text{lin}_{20}$ ($\pm15\%$ of
mag.~$0.05$) for $\gammamm$, followed by $T_{22}$ ($\pm5\%$ of mag.~$0.3$)
for $\gammame$ and check on $\gammamm$, plus $T^\text{lin}_{2,-1}$ ($\pm15\%$ of
mag.~$0.03$) for $\gammaem$.  The different angular dependencies of $T_{21}$
(up to $\pm20\%$ of mag.~$0.08$) can serve as valuable check.

%%%%%%%%%%%%%%%%%%%%%%%%%%%%%%
\subsubsection{Dependence on Rescattering, $\Delta$-Physics and the NN
  Interaction}
\label{sec:moredependences}

As hinted above, reliable theoretical predictions should include a study of
residual theoretical uncertainties. The aforementioned \emph{Mathematica}
notebook therefore explores the influence of NN rescattering, of the dynamical
$\Delta(1232)$, and of the particular two-nucleon interaction used. The
results mostly confirm those of Refs.~\cite{Griesshammer:2010pz,
  Griesshammer:2012we} and thus are only summarised here. Rescattering
significantly affects all observables for energies $\lesssim70\;\MeV$ and is
important to reduce residual dependence on the NN potential and deuteron wave
function up to $120\;\MeV$, as predicted by the power-counting. Details of the
NN potential or deuteron wave function are not reflected in observables. For
example, at $100\;\MeV$, the largest wave-function dependencies are
$\approx\pm5\%$ of the maximum in $T^\text{circ}_{22}$ and $\approx\pm2\%$ of
the maximum in $T^\text{lin}_{20}$.  These observables are however quite small
($<0.05$); all other observables suffer from a residual wave-function
dependence of $<1\%$ at that energy, as tests with AV18~\cite{av18}, Nijmegen
93~\cite{Nijm} and other wave functions demonstrate.

Not surprising is also that $\Delta(1232)$-effects become more pronounced with
increasing energy. It is now well-understood that its spin-flip amplitude
considerably changes the shape of the unpolarised differential cross section
at backward angles~\cite{Hildebrandt:2004hh, Hildebrandt:2005iw,
  Hildebrandt:2005ix}, thereby solving the ``SAL puzzle'' of deuteron Compton
data at $94\;\MeV$~\cite{Hornidge:2000, Levchuk:2000mg, Karakowski:1999pt,
  Karakowski:1999eb, Beane:1999uq, Beane:2002wn, Beane:2004ra}.  While the
influence of the Delta on some observables like $\Sigma^\text{lin}$ may be
very small, it is hard to imagine an EFT without it to be reliable at photon
energies around $100\;\MeV$. As case in point, $T^\text{lin}_{20}$ is at
$100\;\MeV$ increased by $50\%$ and changes shape when the Delta is included;
$T^\text{circ}_{10}$ increases by $30\%$, while $T_{21}$ is reduced by $20\%$,
and $T^\text{lin}_{2,-2}$ even by $50\%$.  $T_{20}$ changes shape at forward
angles.  Delta effects cannot be neglected above about $70\;\MeV$, especially
in the large momentum transfers at back-angles.

%%%%%%%%%%%%%%%%%%%%%%%%%%%%%%%%%%%%%%%%%%%%%%%%%%%%%%%%%%%%%%%%%%%%%%%%%%%%%%%
\section{Conclusions and Outlook}
\setcounter{equation}{0}
\label{sec:conclusions}

Based on a well-known decomposition of the deuteron photo-dissociation cross
section, this work presented a classification of all $18$ independent
observables in Compton scattering off an unpolarised, vector, tensor or
mixed-polarised spin-$1$ target with unpolarised, circularly, linearly or
mixed-polarised beam when final-state polarisations are not detected. The
unpolarised cross section, beam asymmetry, $4$ target asymmetries and $12$
double asymmetries were expressed in terms of the helicity amplitudes and
related to previously used, incomplete parametrisations. This decomposition is
particularly transparent, with each observable readily translated into
specific and well-known beam/target/detector combinations.

The method was then applied to deuteron Compton scattering in \ChiEFT with
dynamical $\Delta(1232)$ degrees of freedom using amplitudes which are
complete at order $e^2\delta^3$ in the energy range from the Thomson limit to
just below the pion production threshold. Since this process tests the
isoscalar two-photon response of the nucleon, embedded in the simplest bound
few-nucleon system~\cite{Griesshammer:2012we}, the sensitivity of each
observable on the $6$ dipole polarisabilities of the nucleon was studied.
These, in turn, encode information on the symmetries and strengths of the
interactions with and between the hadronic internal low-energy degrees of
freedom. They characterise the radiation multipoles which are generated by
displacing the charges and currents inside the nucleon in the electric or
magnetic field of a photon with definite energy and multipolarity. To
determine in particular the $4$ spin polarisabilities is the objective of a
large-scale effort including \HIGS, MAX-Lab and MAMI, since they parametrise
the response of the nucleon spin degrees of freedom but are not yet
well-constrained.  This study thus aids in planning and analysing experiments
to determine the nucleon polarisabilities from deuteron Compton scattering. An
interactive \emph{Mathematica 9.0} notebook of its results over a wide range
of energies is available from \texttt{hgrie@gwu.edu}.

With future high-accuracy determinations of the scalar polarisabilities
$\alphae$ and $\betam$ at lower energies, the spin polarisabilities seem to be
reliably extractable at energies of $\gtrsim100\;\MeV$ from the observables
$T^\text{circ}_{11}$ (circularly-polarised beam on vector target),
$T_{2(2,1)}$ (unpolarised beam on tensor target) and $T^\text{lin}_{2(0,-1)}$
(linearly-polarised beam on tensor target).  This experimentally challenging
but realistic set consists of asymmetries which have maxima from $0.7$ to
$0.05$ and are mostly sensitive to only 1 or 2 polarisabilities. Modifying the
spin polarisabilities by $\pm2\times10^{-4}\;\fm^4$ in them induces variations
of $\pm5\%$ to $\pm20\%$ at $100\;\MeV$.

Since nuclear binding is mediated by charged pion-exchange currents to which
the photons can couple, deuteron Compton scattering concurrently tests the
detailed symmetries and dynamics of the charged part of the two-nucleon
interaction. The $D$ wave contributions of the deuteron wave function and of
the pion-exchange currents lead to nonzero tensor observables. By interference
with the quadrupole components of the incident and outgoing photon, these, in
turn, seem to be much more sensitive on the mixed spin polarisabilities
$\gammaem$ and $\gammame$ than any single-nucleon observable. One may thus
speculate that their determination will first appear from deuteron data -- if
the necessary accuracy can be reached.

\absatz Ongoing work includes embedding the $\calO(e^2\delta^4)$
single-nucleon amplitudes of Ref.~\cite{McGovern:2012ew} for an extension to
photon energies above the pion-production threshold, also with $\Delta$-ful
pion-exchange currents; inclusion of a chirally consistent NN potential; and a
detailed assessment of theoretical uncertainties. In support of ongoing and
planned experiments at \HIGS, MAX-Lab and MAMI, this effort is pursued in the
context of a comprehensive theoretical description of Compton scattering on
the proton, deuteron and \threeHe in \ChiEFT, valid from zero photon energy
well into the $\Delta$ resonance region. As pendant to the present article, a
classification of the independent polarisation transfer observables on a
spin-$1$ target will determine those $5$ which are linearly independent and
complement those presented here for the complete set of $23$ independent
observables~\cite{hgfuture}. From these, the $23$ independent real amplitudes
can be reconstructed in turn, and hence all information accessible in the
two-photon response of the deuteron and its constituents.

Finally, I offer to embed single-nucleon Compton amplitudes, chiral or not,
into the available deuteron code, so that other theoretical
descriptions can be tested collaboratively.

%%%%%%%%%%%%%%%%%%%%%%%%%%%%%%%%%%%%%%%%%%%%%%%%%%%%%%%%%%%%%%%%%%%%%%%%%%%%%%%
%%%%%%%%%%%%%%%%%%%%%%%%%%%%%%%%%%%%%%%%%%%%%%%%%%%%%%%%%%%%%%%%%%%%%%%%%%%%%%%
%%%%%%%%%%%%%%%%%%%%%%%%%%%%%%%%%%%%%%%%%%%%%%%%%%%%%%%%%%%%%%%%%%%%%%%%%%%%%%%

\section*{Acknowledgements}
I am particularly indebted to my Compton collaborators J.~A.~McGovern and
D.~R.~Phillips for discussions and encouragement, and to the organisers and
participants of the INT workshop 12-3: `` Light Nuclei from First Principles''
as well as the ``Workshop to Explore Physics Opportunities with Intense,
Polarized Electron Beams with Energy up to 300 MeV'' at the MIT, both of which
also provided financial support.  D.~G.~Crabb and W.~Meyer shared their
expertise on tensor-polarised targets, and R.~Schiavilla the crucial hint to
Ref.~\cite{Arenhovel:1990yg}. Discussions with W.~Briscoe, E.~Downie,
G.~Feldman and H.~Weller helped clarify various experimental points.  This
work was supported in part by the National Science Foundation under CAREER
award PHY-0645498, by the US-Department of Energy under contract
DE-FG02-95ER-40907, by the Deutsche Forschungsgemeinschaft and the National
Natural Science Foundation of China through funds provided to the Sino-German
CRC 110 ``Symmetries and the Emergence of Structure in QCD'', and by the EPOS
network of the European Community Research Infrastructure Integrating Activity
``Study of Strongly Interacting Matter'' (HadronPhysics3).

\absatz
For Karl Heinz Lindenberger (1925--2012).

% \newpage

% %%%%%%%%%%%%%%%%%%%%%%%%%%%%%%%%%%%%%%%%%%%%%%%%%%%%%%%%%%%%%%%%%%%%%%%%%%%%%%%
% \appendix
% %%%%%%%%%%%%%%% Intro %%%%%%%%%%%%%%%%%%%
% \section{Solving the Integral Equation}
% \setcounter{equation}{0}
% \label{app:appendix}

% %%%%%%%%%%%%%%%
% \subsection{Constructing the Solution}
% \label{app:construction}

% \newpage
%%%%%%%%%%%%%%%%%%%%%%%%%%%%%%%%%%%%%%%%%%%%%%%%%%%%%%%%%%%%%%%%%%%%%%%%%%%%%%%

%%%%%%%%%%%%%%%%%%%%%%%%%%%%%%%%%%%%%%%%%%%%%%%%%%%%%%%%%%%%%%%%%%%%%%%%%%%%%%%%%
%%%%%%%%%%%%%%%%%%%%%%%%%%%%%%%%%%%%%%%%%%%%%%%%%%%%%%%%%%%%%%%%%%%%%%%%%%%%%%%%%
%%%%%%%%%%%%%%%%%%%%%%%%%%%%%%%%%%%%%%%%%%%%%%%%%%%%%%%%%%%%%%%%%%%%%%%%%%%%%%%%%
\newpage\clearpage
\thispagestyle{empty}

\newpage\clearpage
\thispagestyle{empty}

\newpage

\newpage

\newpage

\newpage

\newpage

\newpage\clearpage
\thispagestyle{empty}

\newpage

\newpage

\newpage

\newpage\clearpage

\newpage

\newpage

\newpage\clearpage
\thispagestyle{empty}

\newpage

\newpage

\newpage

\newpage

\end{document}